\documentclass[aps,preprint,preprintnumbers,nofootinbib,showpacs]{revtex4}

\usepackage{amsmath}
\usepackage{amssymb}
\usepackage{mathtools}
\usepackage{dsfont}
\usepackage{color}
\usepackage{graphicx,accents}
\usepackage[normalem]{ulem}

\definecolor{indred}{rgb}{0.8, 0.36, 0.36}
\def\bea{\begin{eqnarray}}
\def\eea{\end{eqnarray}}
\def\sea{\nonumber \\&&}
\def\lla{\left\langle}
\def\rra{\right\rangle}
\def\za{\alpha}
\def\zc{\gamma}
\def\zb{\beta}

\newcommand{\shi}{_{\!\ssc (\!\chi\!)}} 
\newcommand{\shia}{_{\!\ssc (\!\chi\!)A}}
\newcommand{\shib}{_{\!\ssc (\!\chi\!)B}}
\newcommand{\vsi}{_{\!\ssc (\!\varsigma\!)}}

\def\nfac{n_{\!\ssc 0}!\,n_{\!\ssc 1}!\,n_{\!\ssc 2}!\,n_{\!\ssc 3}!}

\def\ssc{\scriptscriptstyle}
\def\lsim{\mathrel{\raise.3ex\hbox{$<$\kern-.75em\lower1ex\hbox{$\sim$}}} }
\def\gsim{\mathrel{\raise.3ex\hbox{$>$\kern-.75em\lower1ex\hbox{$\sim$}}} }
\makeatletter
\DeclareRobustCommand{\cev}[1]{%
  \mathpalette\do@cev{#1}%
}
\newcommand{\do@cev}[2]{%
  \fix@cev{#1}{+}%
  \reflectbox{$\m@th#1\vec{\reflectbox{$\fix@cev{#1}{-}\m@th#1#2\fix@cev{#1}{+}$}}$}%
  \fix@cev{#1}{-}%
}
\newcommand{\fix@cev}[2]{%
  \ifx#1\displaystyle
    \mkern#23mu
  \else
    \ifx#1\textstyle
      \mkern#23mu
    \else
      \ifx#1\scriptstyle
        \mkern#22mu
      \else
        \mkern#22mu
      \fi
    \fi
  \fi
}

\makeatother
\begin{document}
\preprint{{\vbox{\hbox{NCU-HEP-k087}
\hbox{Aug 2020}
\hbox{rev. Dec 2020}
}}}
\vspace*{.7in}


\title{\boldmath Group Theoretical Approach to Pseudo-Hermitian Quantum Mechanics with Lorentz Covariance and $c \to \infty $ Limit
\vspace*{.2in}}

\author{Suzana Bedi\'c,} 

\address{ICRANet, P.le della Repubblica 10, 65100 Pescara, Italy,\\ and
ICRA and University of Rome ``Sapienza'', Physics Department, P.le
A. Moro 5, 00185 Rome, Italy
\vspace*{.3in} }
\author{Otto C. W. Kong, and Hock King Ting}

\address{ Department of Physics and Center for High Energy and High Field Physics,
National Central University, Chung-li, Taiwan 32054  \\
}


\begin{abstract}
\vspace*{.3in}
We present  the formulation of a version
of Lorentz covariant quantum mechanics based on a group
theoretical construction from a Heisenberg-Weyl symmetry
with position and momentum operators transforming 
as Minkowski four-vectors. 
The basic representation is identified as a coherent state
representation, essentially an irreducible component 
of the regular representation, with the matching 
representation of an extension of the group $C^*$-algebra 
giving the algebra of observables. The key feature  
is that it is not unitary but pseudo-unitary, 
exactly in the same sense as the Minkowski spacetime  
representation. The language of pseudo-Hermitian
quantum mechanics is adopted for a clear illustration 
of the aspect, with a metric operator obtained as 
really the manifestation of the Minkowski metric on 
the space of the state vectors. Explicit wavefunction 
description is given without any restriction of the 
variable domains, yet with a finite integral inner product. 
The associated covariant harmonic oscillator Fock state
basis has all the standard properties in exact analog to
those of a harmonic oscillator with Euclidean position 
and momentum operators. Galilean 
limit and the classical limit 
are retrieved rigorously through  appropriate symmetry 
contractions of  the algebra and its representation, 
including the dynamics described through the symmetry 
of the phase space.
\\[.2in]
\end{abstract}

\maketitle

\section{Introduction}
Our group had implemented a, quantum relativity symmetry, 
group theoretical formulation of the full dynamical theory 
of the familiar quantum mechanics with rigorous classical 
limit given as the Newtonian theory, obtained through 
a contraction of the relativity symmetry applied to the 
specific representation \cite{070}. The latter is taken 
as essentially an irreducible component of the regular 
representation of $H(3)$, the Heisenberg-Weyl group. The 
full quantum relativity symmetry, denoted $\tilde{G}(3)$, 
can naturally be seen as a $U(1)$ central extension
 {\cite{u1}} of the Galilean symmetry. $H_{\!\ssc R}(3)$ 
is (or is isomorphic to) its subgroup, left after the 
`time-translation' is taken out. A $H(3)$ representation 
is a spin zero, time independent representation of 
$\tilde{G}(3)$. The representation is really the one 
of the canonical coherent states {\cite{cs1,cs2,cs}}. 
The matching representation of the group $C^*$-algebra
 {\cite{c*1,c*2}}, further extended to a proper class 
of distributions, gives the observable algebra as 
functions, and distributions, of position and momentum 
operators, $\hat{X}_i=x_i\star$ and $\hat{P}_i=p_i\star$, 
as given by the Weyl-Wigner-Groenewold-Moyal(WWGM) 
formulation {\cite{M,w1,w2,w3}}. The operators 
$\za(p_i\star,x_i\star)= \za(p_i,x_i)\star$ act as differential 
operators on the wavefunctions on coherent state basis
$\phi(p^i,x^i)$ by the Moyal star-product $\za\star\phi$; 
$\za\star\zb\star = (\za\star\zb)\star$. $\hat{X}_i$ 
and $\hat{P}_i$ can be seen as operator coordinates
of the quantum phase space \cite{078,081}, which has 
been argued to serve as a proper quantum model for 
the physical space \cite{066,070}. We naturally seek a 
Lorentz covariant version of that with a $c \to \infty$
contraction of the symmetry taking the Lorentz boosts
to that of the Galilean ones \cite{BL}. Such a contraction
is the mathematically rigorous way to look at the 
full approximation of a theory under a certain limit,
from the symmetry theoretical perspective.

The relativity symmetry for the quantum theory is
one of $H_{\!\ssc R}(1,3)$, which fits well into
the contraction chain, at least at the symmetry and 
coset space level  \cite{030,071}. It has been well 
known that from a group theoretical perspective,
a general overcomplete coherent state basis can
naturally be identified with points of the appropriate 
coset. The latter in our cases corresponds to something 
like the classical phase space. The formulation of a fully 
Lorentz covariant version of quantum mechanics, with 
position and momentum operators $\hat{X}_\mu$ and 
$\hat{P}_\mu$ transforming as Minkowski four-vectors, 
has been around since the early days of quantum 
mechanics. A naive thinking would be to 
represent those operators as $x_\mu$ and 
$-i\hbar \partial_{x^\mu}$, respectively,  acting on the 
wavefunctions $\psi(x^\mu)$ with the simple inner 
product giving the squared integral norm, and to take 
a unitary Schr\"odinger evolution under the Einstein 
proper time $\tau$, which gives the Klein-Gordon 
equation as the $\tau$-independent equation of motion. 
Explicit group theoretical picture of that has been 
available since the sixties \cite{Z,J}. The truth is, in any 
theory of quantum mechanics with wavefunctions on 
Minkowski four-vector variable(s), the real symmetry 
behind the system is the $H_{\!\ssc R}(1,3)$ group 
instead of only its Poincar\'e subgroup. There are, 
however, difficulties with the unitary theory, especially 
well illustrated in the covariant harmonic oscillator 
problem \cite{Z,B}, which we show explicitly in the 
appendix. 

Other than being of interest on its own, the
harmonic oscillator problem is of great theoretical
importance. For our usual quantum mechanics, we have
the well appreciated close connection between the Fock
states, as the eigenstates of the harmonic oscillator 
Hamiltonian, and the canonical coherent states. The set 
of Fock states is one of the most useful orthonormal 
basis for the Hilbert space and the latter, as the 
space of rapidly decreasing functions spanned by their 
wavefunctions, giving the states on which the position 
and momentum operators are truly Hermitian in a
completely consistent formulation \cite{M}. Upon  a
more careful inspection, the Fock states are simultaneous
eigenstates of the number operators $\hat{N}_i$, 
or equivalently of $\hat{X}_i^2 +\hat{P}_i^2$ . The 
subspace spanned by the Fock states of a fixed   
eigenvalue $n$ of the total number operator $\sum \hat{N}_i$ 
corresponds exactly to the space of symmetric 
$n$-tensors of the three-dimensional Euclidean space. 
In particular, the three $n=1$ states transform exactly as 
components of a three-vector in a complexified Newtonian 
space. A perfectly nice embedding of all that into the
space spanned by Fock states of the Lorentz covariant 
harmonic oscillator problem should be expected to have 
a fully parallel structure of symmetric $n$-tensors in the 
$(1+3)$-dimensional Minkowski spacetime \cite{B}. Unlike 
for the $SO(3)$ symmetry, however, the noncompact 
nature of Lorentz $SO(1,3)$ symmetry  means that the 
corresponding spaces for the symmetric $n$-tensors, 
as its irreducible representations, cannot be unitary. 

Replacing the full unitarity of the irreducible 
representation of $H_{\!\ssc R}(1,3)$ by a pseudo-unitarity 
exactly in line with the Minkowski spacetime may be 
a good direction to formulate a theory of covariant 
quantum mechanics \cite{082}. The representation 
as one  for the $SO(1,3)$ subgroup would reduce to 
a sum of finite dimensional irreducible components 
each labeled by two integers, the $n$ and a nonzero
positive integer characterizing the spin independent 
Casimir invariant. The latter corresponds to one plus 
the rank of the symmetric $n$-tensors \cite{083}.

We have  presented in Ref.\cite{083} the  complete set 
of Fock states wavefunctions of such a pseudo-unitary  
representation of $H_{\!\ssc R}(1,3)$ symmetry on 
a space of rapidly decreasing functions, hence 
completely free from divergence in themselves as well 
as in the Lorentz invariant indefinite inner product. 
Formulation given there corresponds to writing  the 
time coordinate  as $ict$. Here, the representation is 
rather given in the form of the $\left|p^\mu,x^\mu \rra$ 
coherent states, with $p^\mu$ and $x^\mu$ being real 
Minkowski four-vectors, from our group theoretical  
grand framework \cite{070,030,071}.
 
We want to emphasize that quantum dynamics is 
a symplectic dynamics and the physical Hamiltonian 
is just one among the many general Hamiltonians 
with the generated Hamiltonian flows as symmetries 
of the phase space. It is the symplectic structure of 
the latter as fixed by the inner product, or the metric 
for the vector space or its projective space, that is 
really the key. The actual symmetries of a physical 
system of course correspond to Hamiltonian flows 
the generators of which commute with the physical 
Hamiltonian, with the generators giving the conserved 
physical quantities.  

As a preparation, we first sketch the notion of 
pseudo-Hermiticity and pseudo-unitarity clearly in the
next section. In Sec.\ref{sec2} and \ref{sec3} below, 
we start with an explicit presentation of the regular 
representation, and its irreducible components, of the 
$H(1,3)$ group. A major part of that is also needed 
to formulate the $c \to \infty$ contraction. Each such 
component is shown to give essentially the same physical 
theory of covariant quantum mechanics we present in 
detail on the coherent state basis, in the abstract form 
and in wavefunctions, with the Lorentz invariant indefinite 
inner product. The part that involves the inner product
and the pseudo-Hermitian/pseudo-unitary nature of
the theory is put in Sec.\ref{sec3} after the presentation
of the space of state vectors as the Fock space for 
covariant harmonic oscillator. Sec.\ref{sec4} deals 
with the Lorentz to Galilean, $c \to \infty$, contraction 
of the representation, {\em i.e.} the retrieval of the
`nonrelativistic' limit, the part for the dynamics of which
is left to the last subsection of the Sec.\ref{sec5}. 
The latter is first devoted to the WWGM framework
or the observable algebra, focusing on the symmetry 
transformations and the dynamics as a specific case 
of such a symmetry flow, with the real parameter 
characterizing transformation corresponding to an
evolution parameter, which is taken as the proper 
time in the case. In Sec.\ref{sec6} we give a brief 
description of contraction to the classical theory and conclude in the last section.

\section{Pseudo-Hermiticity and Pseudo-Unitarity}\label{sec1}
Pseudo-unitarity is about an inner product that is not
positive definite, like the Minkowski metric. Lorentz 
covariant quantum theory with an indefinite inner product 
vector space of states was first introduced by Dirac and 
Pauli \cite{D,P}. However, an explicit detailed formulation 
of quantum mechanics with a careful attention paid to the 
covariant and contravariant indices seems not to be available. 
More interest has been focused on quantum field theories, 
such as quantum electrodynamics (see Ref.\cite{N} for 
a review). It has been a common strategy, especially in 
gauge theories since Gupta-Bleuler  \cite{GB1,GB2}, to 
formulate a theory on such a Krein space \cite{Bo} and then 
project it onto the `physical' Hilbert space as the positive 
normed subspace (see also Ref.\cite{B} for the harmonic 
oscillator case), retrieving a standard probability 
interpretation.  Interest in the related subject matter for 
quantum mechanics has been brought back to popularity 
from works on the so-called pseudo-Hermitian quantum 
mechanics \cite{pH1,pH2,pH-M}, which we, in a way, 
rediscovered in our work of Ref.\cite{083}.

Let us sketch  pseudo-Hermitian quantum mechanics 
here. A naive direct picture starts with a Hamiltonian 
operator $\hat{A}_{\!\ssc H}$ that is not Hermitian 
with respect to the given inner product of the Hilbert 
space. If a Hermitian operator $\hat\eta$ can be found 
such that
\begin{equation} \label{pseud}
     \hat{A}_{\!\ssc H}^\dag=\hat\eta  \hat{A}_{\!\ssc H} \hat\eta^{-1} \,,
\end{equation}
the operator $\hat{A}_{\!\ssc H}$ is called
pseudo-Hermitian and $\hat\eta$ the (pseudo-)metric
operator \cite{P,pH-M}. We think metric operator is 
the more appropriate name than pseudo-metric operator,
especially because in our case it is essentially the exact
manifestation of the Minkowski metric $\eta_{\mu\nu}$. 
The interesting thing is that a new inner product 
$ \prescript{}{_\eta}{\!\lla \cdot | \cdot \rra}$ can 
be introduced with respect to which the operator 
$\hat{A}_{\!\ssc H}$ is really Hermitian, namely  
$\hat{A}_{\!\ssc H}^{\dag^\eta}= \hat{A}_{\!\ssc H}$ 
for the Hermitian conjugation satisfying\footnote{
We introduce the somewhat unusual notation for 
a reason. Since we are talking about a second inner
product on the same vector space, we want 
the vectors, kets,  to be independent of the inner 
products, while the sets of bras as functionals can 
be defined differently \cite{083}, giving the different 
Dirac brackets as the different inner products.}
\bea \label{pH}
\prescript{}{_\eta}{\!\lla  \cdot | \hat{A}^{\dag^\eta} \cdot \rra}
= \prescript{}{_\eta}{\!\lla  \hat{A}\cdot | \cdot \rra} \;.
\eea
To be more specific, one can call it a  $\eta$-Hermiticity.

The new inner product is not required to be positive 
definite. More importantly, the evolution generated by 
$\hat{A}_{\!\ssc H}$ is `unitary' \cite{pH-M} in the sense 
that it preserves the inner product between any two 
states. In the case of an indefinite inner product, the 
transformations preserving it are truly represented 
by the pseudo-unitary, rather than unitary, matrices. 
Adopting from the terminology of special relativity,
we have states with norms that can be spacelike (+ve), 
timelike (-ve), or lightlike (0). For the nondegenerate
case, explicitly, one can find a countable orthonormal 
basis, like the Fock basis in our case, with $L$ vectors 
of the norm $-1$, $M$ vectors of $+1$ and none of the 
vanishing norm, in which the `unitary' transformations 
generated by any pseudo-Hermitian operator satisfying 
Eq.(\ref{pH})  are represented by $SU(L,M)$ matrices, 
including the case of $L, M \to \infty$.\footnote{
Ref.\cite{pH-M} restricted the term `inner product'
to positive-definite products, which is not within its
mathematical definition. That is the source of many 
`pseudo-' terminology as in `pseudo-inner product' 
and `pseudo-metric', which  we see as unnecessary.
Defining an absolute pseudo-Hermiticity for the
otherwise Hermitian operators which generate 
pseudo-unitary transformations preserving the
indefinite inner product could be quite sensible though.}
Actually, the theory of quantum mechanics we are
interested in here is a pseudo-unitary representation
of the background (relativity) symmetry group. The 
generators of the symmetry are all pseudo-Hermitian
operators. These are `Hamiltonian operators' in the 
sense of a symplectic/geometric picture  of the theory. 
An acceptable physical Hamiltonian operator in the 
theory, of course, has to satisfy the same 
pseudo-Hermiticity, namely the $\eta$-Hermiticity.

Note that the notion of pseudo-Hermiticity is  
a relative one. $\hat{A}_{\!\ssc H}$ \emph{is not}
Hermitian and \emph{is} pseudo-Hermitian only 
with respect to the original inner product 
$\lla \cdot | \cdot \rra$, for which the Hermitian 
conjugate $\hat{A}_{\!\ssc H}^\dag$ is defined as  
the operator satisfying
\bea
\lla \cdot | \hat{A}^\dag \cdot \rra
= \lla \hat{A}\cdot | \cdot \rra\;.
\eea
Looking at the theory as a dynamical one with 
the physical Hamiltonian operator, the 
$ \prescript{}{_\eta}{\!\lla  \cdot | \cdot \rra}$ 
inner product is the only one relevant. The inner
product certainly gives a metric to the vector 
space and its projective space, which also fixes
the symplectic structure. That is the meaning
of the choice of the (nontrivial) metric operator.
The bottom line is, two different inner products 
on the same vector space really make two 
different inner product spaces and we generally 
do not have any necessity to consider two different
inner products for a single theory of quantum
dynamics. Often time, as in Ref.\cite{083}, it
is just that the simplest or the most familiar kind
of inner product is the `wrong' one, based on 
which one can construct the `right' one more 
easily. In this case, $ \prescript{}{_\eta}{\!\lla  \cdot | \cdot \rra }
  = \lla \cdot | \hat\eta |\cdot \rra$, or
equivalently $ \prescript{}{_\eta}{\!\lla \cdot \right|} = \lla \cdot \right| \hat\eta$.
Obviously, that is the same as 
$\lla \cdot \right|= \prescript{}{_\eta}{\!\lla  \cdot \right|} \hat\eta^{-1}$, 
so the two sets of bras are really on the equal 
footing. The naive perspective that the inner 
product $\lla \cdot | \cdot \rra $ is more basic 
is only a consequence of the presentation.
Furthermore, the reality of an eigenvalue for an 
$\eta$-Hermitian operator follows in the same way 
as for a usual Hermitian one so long as the norm, 
{\em i.e.} $\eta$-norm here, of a corresponding 
eigenstate is nonzero. The latter of course always 
holds on a Hilbert space or, equivalently, for 
a positive definite inner product.

\section{The Irreducible Representations of
\boldmath $H_{\!\ssc R}(1,3)$ \label{sec2}}

We give the Lie algebra for $H_{\!\ssc R}(1,3)$ as 
\bea &&
[J_{\mu\nu}, J_{\rho\sigma}] 
= 2i \left( \eta_{\nu\sigma} J_{\mu\rho} 
  + \eta_{\mu\rho} J_{\nu\sigma} - \eta_{\mu\sigma} J_{\nu\rho} 
   -\eta_{\nu\rho} J_{\mu\sigma}\right) \;,
\sea
[J_{\mu\nu}, Y_\rho] = 2 i  \left( \eta_{\mu\rho} Y_{\nu} 
   - \eta_{\nu\rho} Y
   _{\mu} 
  \right) \;,
  \sea
[J_{\mu\nu}, E_\rho] = 2 i \left( \eta_{\mu\rho} E_{\nu} 
   - \eta_{\nu\rho} E_{\mu} 
  \right) \;,
\sea
[Y_\mu, E_\nu] =2i \eta_{\mu\nu} I\;,
\label{h13}
\eea
where $\eta_{\mu\nu}=\mbox{diag}\{-1,1,1,1\}$. The choice of 
notation with $Y_\mu$ corresponding essentially to spacetime 
position observables and $E_\mu$ to energy-momentum  observables 
is somewhat unusual. The reason for it should be clear from the 
analysis below.  Notice that the generators are all taken to have
no physical dimension, and the factor $2$ corresponds to $\hbar$ 
in the chosen units, which is at least convenient for the coherent 
state formulation \cite{070}. In terms of the group element
$g(p^\mu, x^\mu, \theta,\Lambda^{\!\mu}_\nu)$, 
we have the group product (with the indices suppressed)
\begin{equation}
    g(p', x', \theta', \Lambda') \, g(p, x, \theta,\Lambda)
= g\left(p'+ \Lambda' p, x'+\Lambda' x, 
     \theta'\!+\theta\!- x' \Lambda' p + p'\Lambda' x, 
     \Lambda'\Lambda \right) .\label{group}
\end{equation}
The story is an extension of what has been done in Ref.\cite{066,070} 
for $H_{\!\ssc R}(3)=H\!(3) \rtimes SO(3)$ to the framework of
\bea
H_{\!\ssc R}(1,3)=H\!(1,3) \rtimes SO(1,3) \;,
\eea
the focus of which, for the spin zero case here, is only 
on the irreducible representation of the Heisenberg-Weyl 
symmetry $H\!(1,3)$ and $H\!(3)$. A key point of difference 
between the two cases is that $SO(1,3)$ is noncompact, the 
finite dimensional representations of which, as direct 
extension of those compact ones of $SO(3)$, are pseudo-unitary 
instead of unitary. The basis of that pseudo-unitarity is 
the indefinite Minkowski norm associated with the metric
$\eta_{\mu\nu}$ extending the Euclidean $\delta_{ij}$
 \cite{082,083}. In the case of $H_{\!\ssc R}(3)$, the 
representation is naturally an irreducible component 
of the regular representation of $H\!(3)$, which all 
can be seen actually as physically equivalent. It comes
naturally as wavefunctions in the coherent state basis, 
on which the observables are represented as differential 
operators, essentially  those obtained from the WWGM
framework. Details of all that for the case of $H\!(3)$ 
group has been presented in Ref.\cite{070}.

We  present first the results from a harmonic analysis 
of  Heisenberg-Weyl groups  adapted to our case of 
$H(1,3)$ \cite{T}. We write the left regular 
representation  in $\hbar=2$ units as 
$V (p^\mu,x^\mu,\theta)
   =e^{i(p^\mu Y^{\!\ssc L}_\mu -x^\mu E^{\!\ssc L}_\mu+\theta I^{\!\ssc L})}$,
where 
\bea
Y^{\!\ssc L}_\mu &=&i  x_\mu  \partial_\theta +  i \partial_{p^\mu}\;,
\nonumber \\
E^{\!\ssc L}_\mu &=& i  p_\mu \partial_\theta -  i \partial_{x^\mu}\;,
\nonumber \\
I^{\!\ssc L} &=& i \partial_\theta \;
\eea
are the left-invariant vector fields. In an irreducible 
representation,  the central generator $I$ has  to be 
represented by a multiple of identity. We write the one 
parameter series $V_{\!\ssc\varsigma}$ ($\varsigma\ne 0$)
of representations for the generators as operators 
$\{\hat{Y}_{\!\ssc \varsigma}^{\!\ssc L}, \hat{E}_{\!\ssc \varsigma}^{\!\ssc L}, \varsigma \hat{I}\}$, 
where $\hat{I}$ is the identity operator and 
$[\hat{Y}^{\!\ssc L}_{\!\ssc \varsigma \mu}, \hat{E}^{\!\ssc L}_{\!\ssc \varsigma \nu}]
   =2 i\varsigma\eta_{\mu\nu} \hat{I}$.
The ${V}^{\!\ssc L}_{\!\ssc\varsigma}$ set can be considered 
the set of equivalence classes of irreducible representations
with nonzero Plancherel measure. The limit of 
$V^{\!\ssc L}_{\!\ssc \varsigma}$ as $\varsigma \to 0$ 
gives the whole set of irreducible one-dimensional 
representations. The latter set has zero Plancherel measure and
together with the $V^{\!\ssc L}_{\!\ssc \varsigma}$ exhausts all
equivalence classes of irreducible representations. Based on the 
measure, one should consider the expansion 
\bea \label{za}
\za(p^\mu, x^\mu,\theta)=  \frac{1}{(2\pi)^{\frac{1}{2}}}\int \!\! d\varsigma \, 
   \za_{\ssc\varsigma} (p^\mu, x^\mu) \,e^{-i\varsigma\theta} |\varsigma|^n\;,
\eea
$n=1+3$ here, given as the inverse Fourier-Plancherel  transform. 
The actions of the left-invariant vector fields on $\za(p,x,\theta)$ 
in the form of Eq.(\ref{za}) are given by their actions on 
$\za_{\ssc \varsigma}(p,x)e^{-i\varsigma\theta}$ parts as 
${\varsigma}x+i \partial_{p}$, ${\varsigma}p-i \partial_{x}$, 
and $\varsigma$, respectively.  Here, and below, we suppress 
the indices wherever it is unambiguous. We can see that the 
action at each $\varsigma \ne 0$ corresponds exactly to the  
${V}^{\!\ssc L}_{\!\ssc\varsigma}$ representation 
with the generators represented by  $\{ \hat{Y}^{\!\ssc L}_{\!\ssc \varsigma}, 
  \hat{E}^{\!\ssc L}_{\!\ssc \varsigma},  \varsigma \hat{I}\}$. 
That is the reduction of the regular representation into 
irreducible components. For positive values of  $\varsigma$, 
one can introduce the $\varsigma$-independent operators
\bea && \label{genop}
\hat{X}^{\!\ssc L}_{\!\ssc (\!\varsigma\!)\mu} \equiv \frac{1}{\sqrt{\varsigma}} \hat{Y}^{\!\ssc L}_{\ssc \!\varsigma \mu}
  = x_{\ssc (\!\varsigma\!)\mu}  +  i \partial_{p^\mu_{\ssc (\!\varsigma\!)}}\;,
\sea
\hat{P}^{\!\ssc L}_{\!\ssc (\!\varsigma\!)\mu} \equiv \frac{1}{\sqrt{\varsigma}} \hat{E}^{\!\ssc L}_{\ssc \!\varsigma \mu}
  = p_{\ssc (\!\varsigma\!)\mu}  -  i \partial_{x^\mu_{\ssc (\!\varsigma\!)}}\;,
\eea
where we have $x_{\ssc (\!\varsigma\!)}= \sqrt{\varsigma} x$ 
and $p_{\ssc (\!\varsigma\!)}= \sqrt{\varsigma} p$. 
$V^{\!\ssc L}_{\varsigma}(p_{\ssc (\!\varsigma\!)},   x_{\ssc (\!\varsigma\!)},
  \theta_{\ssc (\!\varsigma\!)})$ is then given by
$e^{i(p_{\ssc (\!\varsigma\!)}\hat{X}^{\!\ssc L}_{\!\ssc (\!\varsigma\!)}
   -x_{\ssc (\!\varsigma\!)}\hat{P}^{\!\ssc L}_{\!\ssc (\!\varsigma\!)}
    +\theta_{\ssc (\!\varsigma\!)} \hat{I})}$,
with $\theta_{\ssc (\!\varsigma\!)}={\varsigma} {\theta}$, 
hence in a form formally independent of $\varsigma$. 
$\hat{X}^{\!\ssc L}_{\!\ssc (\!\varsigma\!)}$ and
 $\hat{P}^{\!\ssc L}_{\!\ssc (\!\varsigma\!)}$ are still $SO(1,3)$ 
vectors, and so are $p_{\ssc (\!\varsigma\!)}$ and 
$x_{\ssc (\!\varsigma\!)}$. The $(\!\varsigma\!)$ index 
becomes completely dummy and analysis based on the new 
operators and  parameters is independent of $\varsigma$ so long 
as we are looking only at a particular irreducible representation.  
One can even simply drop it. From a physics perspective, we 
have absorbed the value of $\varsigma$ by a choice of 
physical unit for measuring the observables corresponding 
to $Y$ and $E$, here all in the unit of  $\sqrt{\varsigma}$. 
 For $\varsigma$ being negative\footnote{
From the physical point of view, the representations corresponding 
to different values of $\varsigma$ can be seen as describing the 
same physics.  The parameter $\varsigma$ may then be taken as 
the physical constant $\frac{\hbar c^2}{2}$.  And for that matter, 
$\varsigma$ cannot be negative. Physicists identify the symmetry 
algebra from a relevant representation with 
$\hat{X}^{\!\ssc L}_{\!\ssc (\!\varsigma\!)}$ and 
$\hat{P}^{\!\ssc L}_{\!\ssc (\!\varsigma\!)}$ 
as the position and momentum observables satisfying
$[\hat{X}^{\!\ssc L}_{\!\ssc (\!\varsigma\!)\mu},\hat{P}^{\!\ssc L}_{\!\ssc (\!\varsigma\!)\nu}]
=2i\eta_{\mu\nu}$,  in the $\hbar=2$ units. However, 
the mathematical case of a product of two 
representations with different $\varsigma$ values 
may have interesting physics implications if  a composite 
physical system corresponding to that exists in nature.  
},
 we should switch $\hat{Y}^{\!\ssc L}_{\!\ssc \varsigma}$ 
with $\hat{E}^{\!\ssc L}_{\!\ssc \varsigma}$ first;   
{\em i.e.} we take  
\bea &&
\hat{X}^{\!\ssc L}_{\!\ssc (\!\varsigma\!)} 
           \equiv \frac{1}{\sqrt{|\varsigma|}} \hat{E}^{\!\ssc L}_{\ssc \!\varsigma}
  = x_{\ssc (\!\varsigma\!)}  +  i \partial_{p_{\ssc (\!\varsigma\!)}}\;,
\sea\nonumber
\hat{P}^{\!\ssc L}_{\!\ssc (\!\varsigma\!)} 
              \equiv \frac{1}{\sqrt{|\varsigma|}} \hat{Y}^{\!\ssc L}_{\ssc \!\varsigma}
  = p_{\ssc (\!\varsigma\!)}  -  i \partial_{x_{\ssc (\!\varsigma\!)}}\;,
\eea
achieved by taking $x_{\ssc (\!\varsigma\!)}= -\sqrt{|\varsigma|} p$ 
and $p_{\ssc (\!\varsigma\!)}= -\sqrt{|\varsigma|} x$.  
The result still maintains
$V^{\!\ssc L}_{\varsigma}(p_{\ssc (\!\varsigma\!)},   x_{\ssc (\!\varsigma\!)},
  \theta_{\ssc (\!\varsigma\!)})$.
$\varsigma$ can actually be seen as the eigenvalue of $I$, 
essentially the Casimir operator. The semi-direct product 
structure $H_{\!\ssc R}(1,3)=H\!(1,3) \rtimes SO(1,3)$ says 
that with each irreducible  representation of the subgroup 
$H\!(1,3) \rtimes S_{\hat{O}}$, where $S_{\hat{O}} \subseteq SO(1,3)$ 
is the stability subgroup for an orbit $\hat{O}$ of $SO(1,3)$ 
in the space of equivalent classes of irreducible  representations 
of $H(1,3)$, one can associate an induced representation which 
is irreducible \cite{BR}.  We have seen that, apart from the set 
of measure zero, each   of which only gives one-dimensional 
representations, the irreducible representations are 
characterized by the nonzero value of $\varsigma$ and the
representations (though mathematically nonequivalent) 
can be cast in the same form as
$V^{\!\ssc L}_{\varsigma}(p_{\ssc (\!\varsigma\!)},   x_{\ssc (\!\varsigma\!)},
  \theta_{\ssc (\!\varsigma\!)})$.
It is obvious that the representation is invariant under the $SO(1,3)$
transformations, hence each is an independent orbit. That is to say 
$S_{\hat{O}} = SO(1,3)$. The fact is of paramount importance for
unambiguously identifying the nature of the coherent states below. 
In view of the discussion above, we can see that for any of the 
$V^{\!\ssc L}_{\varsigma}(p_{\!\ssc (\!\varsigma\!)},   x_{\ssc (\!\varsigma\!)},
  \theta_{\!\ssc (\!\varsigma\!)})$
representation, we can simply write it in the simple notation
$V^{\ssc\! L}\!(p,x,\theta)$, like taking the $\varsigma=1$ 
case as a representative. That is essentially what has been 
done in Ref.\cite{070} for the $H\!(3)$ or $H_{\!\ssc R}(3)$ 
case. However, for the reason to be clear below, 
we keep the explicit $\varsigma$-notation for the 
most part of the manuscript.

The standard approach is to introduce the abstract 
canonical coherent states as
\bea \label{CSdef}
\left|p^\mu\vsi,x^\mu\vsi \rra
\equiv V_{\ssc \!\varsigma}\!(p^\mu_{\!\ssc (\!\varsigma\!)},x^\mu_{\ssc (\!\varsigma\!)})
       \left|0,0 \rra
 \equiv e^{-i\theta_{\ssc (\!\varsigma\!)}} 
 {V}_{\ssc \!\varsigma}(p^\mu_{\!\ssc (\!\varsigma\!)},x^\mu_{\ssc (\!\varsigma\!)},
     \theta_{\!\ssc (\!\varsigma\!)})\left|0,0 \rra  \;, 
\eea
for
\bea
V_{\ssc \!\varsigma}(p^\mu_{\!\ssc (\!\varsigma\!)},x^\mu_{\ssc (\!\varsigma\!)},\theta_{\ssc (\!\varsigma\!)})
 \equiv e^{i (p^\mu_{\!\ssc (\!\varsigma\!)}\hat{X}_{\ssc (\!\varsigma\!)\mu}
  - x^\mu_{\ssc (\!\varsigma\!)}\hat{P}_{\ssc (\!\varsigma\!)\mu} 
     +\theta_{\!\ssc (\!\varsigma\!)}\hat{I})} \;,
\eea
representing the $H(1,3)$  group element
${W}\!(p^\mu\vsi, x^\mu\vsi, \theta\vsi)$ 
satisfying the group product
\bea\label{wwt}
{W}\!(p'^\mu\vsi, x'^\mu\vsi, \theta'\vsi)  {W}\!(p^\mu\vsi, x^\mu\vsi, \theta\vsi)
= {W}\!\!\left(p'^\mu\vsi + p^\mu\vsi, x'^\mu\vsi + x^\mu\vsi, 
     \theta'\vsi+\theta\vsi- \!(x'_{{\ssc (\!\varsigma\!)}_\mu}  p^\mu\vsi \!- p'_{{\ssc (\!\varsigma\!)}_\mu} x^\mu\vsi) \right) .
\eea 
Each group element can be identified with 
a point in the  $H_{\!\ssc R}(1,3)/SO(1,3)$ coset 
space \cite{066,071}. $\hat{X}_{\ssc (\!\varsigma\!)}$ 
and $\hat{P}_{\ssc (\!\varsigma\!)}$ are 
operators on the abstract representation space  
$\mathcal{H}_{\ssc \varsigma}$ spanned by the
$\left|p^\mu_{\ssc (\!\varsigma\!)},x^\mu_{\ssc (\!\varsigma\!)} \rra$  
vectors, and
$\left|0,0 \rra = \left|0_{\ssc (\!\varsigma\!)} \rra $ 
is a fiducial normalized cyclic vector corresponding 
to the points $(0,0,\theta_{\!\ssc (\!\varsigma\!)})$
in the coset space, each of which is fixed under 
$SO(1,3)$ transformations.

\section{Pseudo-Hermitian Nature of the
Representation of Symmetry Generators from the Fock States} \label{sec3}
The kind of operator representation of the four-vector observables 
given in Eq.\eqref{genop} would be naively seen as Hermitian,
hence the full representation of the $H_{\!\ssc R}(1,3)$ group
as unitary. To be more careful, the unitarity of a representation
is really to be defined with respect to the inner product assumed
for the representation space. The operator representation sure
is Hermitian with respect to the usual squared-integral inner
product, (with bar denoting the complex conjugation),
\begin{equation} \label{unitinner}
 \lla \phi | \phi' \rra  =\frac{1}{\pi^4}  
 \int\!\! d^4p \,d^4x \; 
     \bar{\phi}(p^\mu,
     x^\mu) \, {\phi'}\!(p^\mu,x^\mu) \;,
\end{equation}
for the wavefunctions that vanish at infinity. In the 
equivalent formulation in terms of standard Schr\"odinger
wavefunctions $\psi(x^\mu)$, that is exactly the unitary
representation given explicitly first in 1966 \cite{Z,J}. 
The short-comings  of the formulation are best seen 
in the covariant harmonic oscillator problem \cite{Z,B}. 
We illustrate them explicitly in  the  appendix.
To illuminate the pseudo-Hermitian nature of our 
representation, we present in the following the 
pseudo-unitary Fock space and complete the coherent 
states representation, together with the appropriate 
Lorentz invariant integral inner product. For convenience,
in this section we drop the $_{\!\ssc (\!\varsigma\!)}$ 
and $_{\!\ssc \varsigma}$ subscripts.

We start with summarizing the more transparent abstract 
algebraic results for the Fock space \cite{B,083} 
in a better logic and   notation. For the Hamiltonian 
$\hat{X}_\mu \hat{X}^\mu + \hat{P}_\mu \hat{P}^\mu$, 
we consider $\hat{a}^\mu=\eta^{\mu\nu} (\hat{X}_\nu+i\hat{P}_\nu)$,
$\hat{a}^{\dag^\eta}_\mu=\hat{X}_\mu-i\hat{P}_\mu$
and $\hat{N}_{\!\ssc(\mu)}=\frac{1}{4}\hat{a}_\mu^{\dag^\eta}  \hat{a}^\mu$
without summation (the index $(\mu)$ is not a vector one) with
\bea\label{Naa}
[\hat{N}_{(\mu)},\hat{a}^\nu] = - \eta_\mu^\nu \hat{a}^\mu\;,
\qquad  
[\hat{N}_{(\mu)},\hat{a}^{\dag^\eta}_\nu] = \eta^\mu_\nu \hat{a}^{\dag^\eta}_\mu \;;
\qquad
\left[\hat{a}^\mu,\hat{a}^{\dag^\eta} _\nu \right]=4\eta^{\mu}_\nu\;.
\eea
Here, we introduce the $^{\dag^\eta}$ notation for a yet
unspecified $\hat{\eta}$, not excluding that being trivial,
requiring however the $\eta$-Hermiticity of $\hat{X}_\mu$ 
and $\hat{P}_\mu$. Note that the feasible inner product 
is still to be determined. The Fock states are  simultaneous
eigenstates of the  $\hat{N}_{\!\ssc(\mu)}$, and hence 
also the $\hat{N}=\sum \hat{N}_{\!\ssc(\mu)}$, operators.
The last one is of course Lorentz invariant.
\begin{equation} 
   \left|n\rra \equiv
    \left| n_{\! \ssc 0};n_{\! \ssc 1},n_{\! \ssc 2},n_{\! \ssc 3}\right\rangle  
=   \frac{1}{2^{n} \sqrt{\nfac}} \left(\hat{a}_{\!\ssc 0}^{\dag^\eta} \right)^{n_{0}} 
   \left(\hat{a}_{\!\ssc 1}^{\dag^\eta} \right)^{n_1} 
     \left(\hat{a}_{\!\ssc 2}^{\dag^\eta}\right)^{n_2}
  \left(\hat{a}_{\!\ssc 3}^{\dag^\eta} \right)^{n_3}
    |0\rangle \;,
\end{equation}
with
\bea &&
\hat{a}^{\ssc 0}    \left| n_{\! \ssc 0};n_{\! \ssc 1},n_{\! \ssc 2},n_{\! \ssc 3} \rra
  =  2 \sqrt{n_{\! \ssc 0}}  \left| n_{\! \ssc 0}-1;n_{\! \ssc 1},n_{\! \ssc 2},n_{\! \ssc 3} \rra \;,
\sea
\hat{a}_{\ssc 0}^{\dag^\eta} \left| n_{\! \ssc 0};n_{\! \ssc 1},n_{\! \ssc 2},n_{\! \ssc 3} \rra
  =  2 \sqrt{(n_{\! \ssc 0}+1)}   \left| n_{\! \ssc 0}+1;n_{\! \ssc 1},n_{\! \ssc 2},n_{\! \ssc 3} \rra  \;,
\label{ud}
\eea
and the exact corresponding results for $\hat{a}^i \left| n  \rra$
and $\hat{a}_i^{\dag^\eta} \!\left| n  \rra$. The $H_{\!\ssc R}(1,3)$ 
canonical coherent states, satisfying $\hat{a}^\nu \left|p^\mu,x^\mu\rra
       = 2 \left(x^{\nu}+ip^{\nu} \right) \left|p^\mu,x^\mu\rra$,  
can be expanded as 
\bea \label{CSn}
\left|p^\mu,x^\mu\rra
 &=& e^{-\frac{x_\mu x^\mu+ p_\mu p^\mu}{2}}
   \sum\frac{1}{\sqrt{n_{\! \ssc 0}!\, n_{\! \ssc 1}!\, n_{\! \ssc 2}! \, n_{\! \ssc 3}! }}
   \left(x^{\ssc 0}+ip^{\ssc 0} \right)^{n_0}
   \left(x^{\ssc 1}+ip^{\ssc 1}\right)^{n_1} 
\sea \qquad \qquad
   \times \left(x^{\ssc 2}+ip^{\ssc 2} \right)^{n_2} 
   \left(x^{\ssc 3}+ip^{\ssc 3} \right)^{n_3} \,
   \left| n_{\! \ssc 0};n_{\! \ssc 1},n_{\! \ssc 2},n_{\! \ssc 3} \rra \;,
\eea
with $\left|0,0\rra= \left|0\rra$. Moreover, those are  
exactly the states defined earlier, obtained by an action of
 $V\!(p^\mu,x^\mu) = e^{i(p^\mu \hat{X}_\mu - x^\mu \hat{P}_\mu )}$ 
on $\left|0\rra$ state, or equivalently
\[
V\!(p^\mu,x^\mu) \left|0\rra
  = e^{-\frac{x_\mu x^\mu+ p_\mu p^\mu}{2}} 
      e^{ \frac{  (x^{\nu}+ip^{\nu}) \hat{a}_{\nu}^{\dag^\eta} }{2} }
    |0\rangle \;.
\] 

The right inner product to complete the familiar algebra 
of the problem is, however, nontrivial. While the operators 
$\hat{a}^\mu$, $\hat{a}^{\dag^\eta}_\mu$ and $\hat{N}_{\!\ssc(\mu)}$ 
with the commutation relation of Eq.(\ref{Naa}) give 
a convenient generalization of the $\hat{a}_i\equiv \hat{a}^i$, 
$\hat{a}^{\dag}_i$ and $\hat{N}_{\!\ssc(i)}$ system, 
insensitive to the metric signature yet having the right Lorentz
transformation properties of the Fock state solutions built 
in, $\hat{a}^{\dag^\eta}_{\ssc 0}$ is not a naive Hermitian
conjugate of $\hat{a}_{\ssc 0}$. In fact, we can obtain
from Eq.(\ref{ud}) that $\lla \hat{a}^\mu\cdot |  \cdot \rra
   = \lla \cdot |\hat{a}_\mu^{\dag^\eta}  \cdot \rra$ when 
the usual orthonormality $\lla m | n \rra=\delta_{mn}$ is
assumed. We need a new inner product defined as
\bea
 \prescript{}{_\eta}{\!\lla  m | n \rra} = (-1)^{n_0} \,\delta_{mn} \;,
\eea
{\em i.e.} $\hat\eta = \sum \left(-1)^{n_0} |n \rra \! \lla n \right|$,
with the corresponding $\eta$-Hermitian conjugation,  
$\hat{a}^{\dag^\eta}_\mu=\hat\eta^{-1} \hat{a}^{\dag}_\mu \hat\eta$,  
giving $ \prescript{}{_\eta}{\!\lla  \hat{a}_\mu \cdot |  \cdot \rra}
   =  \prescript{}{_\eta}{\!\lla  \cdot |\hat{a}^{\dag^\eta}_\mu \cdot \rra}$. 
Note that specifying the inner product for a complete basis
uniquely defines the inner product over the whole space. 
One can easily see that 
$\left(\hat{a}^\mu\right)^\dag= \hat{a}^{\dag^\eta}_\mu$ 
implies Hermiticity of  $\hat{X}_i  $ and $\hat{P}_i$ 
operators, while from equating 
$\left(\hat{a}^0\right)^{\!\dag}=\left(\hat{X}^{\ssc 0}+i\hat{P}^{\ssc 0}\right)^{\!\!\dag}$ 
with $ \hat{a}^{\dag^\eta}_0$, we see that $\hat{X}_{\ssc 0}$ 
and $\hat{P}_{\ssc 0}$ are anti-Hermitian with respect
to the usual inner product, {\em i.e.}  the one with 
$\hat{\eta}$ being the identity.  However, all of the 
$H_{\!\ssc R}(1,3)$ generators are represented by 
pseudo-Hermitian, or $\eta$-Hermitian, operators. 
The operators all have real spectra, as we  show 
explicitly below. Explicitly, $\hat{X}_\mu$, $\hat{P}_\mu$ 
and $\hat{J}_{\mu\nu}= \hat{X}_\mu \hat{P}_\nu   
   -  \hat{X}_\nu \hat{P}_\mu$ (and $\hat{I}$) all 
satisfy Eq.(\ref{pseud}) in the place of $\hat{A}_{\!\ssc H}$,
and we have a pseudo-unitary representation with invariant 
inner product $ \prescript{}{_\eta}{\!\lla  \cdot |\cdot \rra} $. 
In particular, the coherent states are normalized as 
$ \prescript{}{_\eta}{\!\lla  0 |0 \rra}
   = \prescript{}{_\eta}{\!\lla  p^\mu,x^\mu |p^\mu,x^\mu\rra }=1$, 
hiding the inner product indefiniteness. We can see the 
latter  either through the explicit use of Eq.\eqref{CSn}, 
or directly from the fact that $V\!(p^\mu,x^\mu)$ 
in Eq.\eqref{CSdef} is an $\eta$-unitary operator. 
From the definition of  $\hat{\eta}$ in Fock basis and 
Eq.\eqref{CSn} we obtain
\begin{equation}
    \prescript{}{_\eta}{\!\lla p^\mu,x^\mu\right|}
  = \lla p^\mu,x^\mu \right| \hat{\eta}
  =\lla p_{\mu},x_{\mu} \right| \;,
\end{equation}
showing explicitly the metric operator $\hat\eta$ is
a direct manifestation of the Minkowski metric in the 
Krein space of our quantum theory, exactly as we 
are looking for \cite{082}.

As the state $\left|0,0\rra=\left|0\rra$ has zero expectation values 
for the $\hat{X}_\mu$ and $\hat{P}_\mu$  operators, we get 
\bea &&
\prescript{}{_\eta}{\!\lla  p^\mu,x^\mu| \hat{X}_\nu |p^\mu,x^\mu \rra }
=2x_\nu \;,
\sea
\prescript{}{_\eta}{\!\lla  p^\mu,x^\mu| \hat{P}_\nu |p^\mu,x^\mu \rra }
=2p_\nu \;.
\eea
The generic wavefunctions can  be introduced as  
$\phi(p^\mu,x^\mu) \equiv  \prescript{}{_\eta}{\!\lla  p^\mu,x^\mu| \phi\rra}$, 
satisfying
\bea 
 \prescript{}{_\eta}{\!\lla  p^\mu,x^\mu  \left| \hat{X}_{\nu} \right|\phi \rra }
  &=& \hat{X}^{\!\ssc L}_{\nu}   \phi (p^\mu,x^\mu)  \;,
\nonumber \\
 \prescript{}{_\eta}{\!\lla p^\mu,x^\mu  \left| \hat{P}_{\nu} \right|\phi \rra} 
   &=& \hat{P}^{\!\ssc L}_{\nu}   \phi (p^\mu,x^\mu) \;,
\eea
with
\bea
\hat{X}^{\!\ssc L}_{\mu} &=&    x_{\mu}  +  i \partial_{\!p^\mu }\;,
\nonumber \\
\hat{P}^{\!\ssc L}_{\mu} &=&   p_{\mu}  - i \partial_{\!x^\mu }\;,
\label{L}
\eea
exactly in the form of Eq.\eqref{genop} and
\bea \label{v-shift} &&
{V}^{\ssc\! L}\!(p^\mu,x^\mu) 
  \phi(p'^\mu,x'^\mu)   \equiv
 \prescript{}{_\eta}{\!\lla  p'^\mu,x'^\mu 
 \left|{V}(p^\mu,x^\mu)  \right|\phi\rra }
\sea\hspace*{.5in}
= \phi (p'^\mu- p^\mu,
 x'^\mu-x^\mu)
  e^{{i}(x'_{\mu} p^\mu - p'_{\mu} x^\mu )}\;.
\eea
We see that the abstract formulation from the set 
of canonical coherent states based on the $H\!(1,3)$ 
manifold and the one from the irreducible component 
of the regular representation are really the same one. 
Wavefunction of a coherent state labeled by $A$ is given by
\bea\label{molap}
\phi_{\!\ssc A} (p^\mu,x^\mu)\equiv
  \prescript{}{_\eta}{\!\lla  p^\mu,x^\mu| p^\mu_{\!\ssc A}, x^\mu_{\ssc A}\rra}
=e^{i\left( x_{\mu} p^\mu_{\!\ssc A}-p_{\mu} x^\mu_{\!\ssc A} \right) }
e^{-\frac{1}{2}\left[\left(x-x_{\!\ssc A} \right)^2+\left(p-p_{\!\ssc A}   \right)^2 \right]} \;,
\eea
where $\left(x-x_{\!\ssc A}  \right)^2$ and
$\left(p-p_{\!\ssc A}   \right)^2$ are the Minkowski 
vector magnitude squares, and can be seen as 
a special case of Eq.\eqref{v-shift}, namely
\begin{equation}
    \phi_{\!\ssc A}(p^\mu,x^\mu) 
= \prescript{}{_\eta}{\!\lla  p^\mu,x^\mu 
 \left|{V}\!( p^\mu_{\!\ssc A},x^\mu_{\!\ssc A}) \right|0,0\rra }
= {V}^{\ssc\! L}\!( p^\mu_{\!\ssc A},x^\mu_{\ssc A}) \phi_o (p^\mu,x^\mu)\;.
\end{equation}
In particular, we have
\[
\phi_o(p^\mu,x^\mu) 
  = \prescript{}{_\eta}{\!\lla p,x  |0\rra}
=e^{-\frac{p_{\mu} p^\mu +x_{\mu}x^\mu}{2}}\;,
\]
which is the Lorentz invariant symmetric Gaussian.

To obtain the inner product on the space of wavefunctions, 
one simply has to look for the proper resolution of the identity
operator on the Krein space. We have 
\bea
\hat{I} &=&  \sum \left|n\rra\! \prescript{}{_\eta}{\!\lla  n \right|}  \hat\eta
  = \int \!\! {d^3\!p d^3\!x dp^{\ssc 0}  dx^{\ssc 0}}
    \frac{e^{-2\left(x^{\ssc 0}\!\right)^{\!2}-2\left(p^{\ssc 0}\!\right)^{\!2} }  }{\pi^4}
   \left|p^\mu,x^\mu\rra \! \prescript{}{_\eta}{\!\lla  p^\mu,x^\mu \right|}
    \hat\eta\;.
\eea
Therefore, the functional  $ \prescript{}{_\eta}{\!\lla  \psi \right|}$
is represented on the space of $\phi (p^\mu,x^\mu)$ as
\[
\int \!\! {d^3\!p d^3\!x dp^{\ssc 0}  dx^{\ssc 0}}
    \frac{e^{-2\left(x^{\ssc 0}\!\right)^{\!2}-2\left(p^{\ssc 0}\!\right)^{\!2} }  }{\pi^4}
   \bar{\psi}\! (p^i,x^i, -p^{\ssc 0},-x^{\ssc 0}) \,\bigg(\cdot\bigg) \;,
\]
with the very nontrivial integration measure.
The inner product $ \prescript{}{_\eta}{\!\lla \psi | \phi \rra}$ 
is then given by
\bea \label{ptip}
 \prescript{}{_\eta}{\!\lla \psi | \phi \rra}= \frac{1}{\pi^4}\int \!\! {d^3\!p d^3\!x dp^{\ssc 0}  dx^{\ssc 0}}
   \frac{\bar{\psi}\! (p^i,x^i, -p^{\ssc 0},-x^{\ssc 0}) }{e^{\left(x^{\ssc 0}\!\right)^{\!2}+\left(p^{\ssc 0}\!\right)^{\!2} }  } 
  \frac{\phi (p^\mu,x^\mu)}{e^{\left(x^{\ssc 0}\!\right)^{\!2}+\left(p^{\ssc 0}\!\right)^{\!2} } }\;.
\eea
Each of the basis functions ${\phi_n (p^\mu,x^\mu)}$,
and hence any general ${\phi (p^\mu,x^\mu)}$ 
in the spanned space, is formally divergent at timelike 
infinity of the four-vector variables. On the other hand, all
$\frac{\phi_n (p^\mu,x^\mu)}{e^{\left(x^{\ssc 0}\!\right)^{\!2}+\left(p^{\ssc 0}\!\right)^{\!2} } }$,
and hence all 
$\frac{\phi (p^\mu,x^\mu)}{e^{\left(x^{\ssc 0}\!\right)^{\!2}+\left(p^{\ssc 0}\!\right)^{\!2} } }$,
are rapidly decreasing functions. 
The factor ${e^{-\left(x^{\ssc 0}\!\right)^{\!2}-\left(p^{\ssc 0}\!\right)^{\!2} } }$
 takes the ${e^\frac{{\left(x^{\ssc 0}\!\right)^{\!2}+\left(p^{\ssc 0}\!\right)^{\!2} } }{2} }$
factor in all ${\phi_n (p^\mu,x^\mu)}$ back to
${e^-\frac{{\left(x^{\ssc 0}\!\right)^{\!2}+\left(p^{\ssc 0}\!\right)^{\!2} } }{2} }$, 
which characterizes the class of functions. The integral is
finite for all wavefunctions as finite linear combinations
of the Fock state basis $\phi_n$.  Using
$\frac{\phi (p^\mu,x^\mu)}{e^{\left(x^{\ssc 0}\!\right)^{\!2}+\left(p^{\ssc 0}\!\right)^{\!2} } }$
as the wavefunctions cannot be correct, though. That 
would, for example, make the wavefunction for 
$\left|0 \rra$ not Lorentz invariant and mess up the 
right transformation properties of all those for the 
Fock states, described in Ref.\cite{083}. Thinking 
further about ${\psi^*\! (p^i,x^i, -p^{\ssc 0},-x^{\ssc 0}) }$
as ${\psi^*\! (p_\mu,x_{\mu}) }$,
one can see in hindsight that the inner 
product expression is indeed exactly what it should be.
Of course we have that here rigorously established.

One can now easily show that the Fock states wavefunctions 
$\phi_n (p^\mu,x^\mu)$  have the proper norm  $\pm 1$, 
and therefore are non-divergent without restricting the 
domain. The analytical feature is much better than that
of the unitary representation (see the appendix). 
Note that, other than having a different inner product with
a nontrivial integration measure, our formulation in 
terms of the wavefunctions and differential operator
representation of the $\hat{X}_\mu$ and $\hat{P}_\mu$
really are the same as the usual unitary one. 
That illustrates clearly that the basic observables 
$\hat{X}_\mu$ and $\hat{P}_\mu$, as well as other
observables in the form of their real functions, all have
the same eigenvalues and eigenfunctions. In particular,
the spectra are real.

\section{Lorentz to Galilean Contraction} \label{sec4}
A contraction {\cite{G,IW}} of the Lorentz symmetry $SO(1,3)$, 
sitting inside the $H_{\!\ssc R}(1,3)$, to the Galilean 
$ISO(3)$ has been discussed in Ref.\cite{071}, together with 
the corresponding coset spaces of interest. The full (quantum)
relativity symmetry group obtained by contraction is named 
$H_{\!\ssc GH}(3)$, with commutators among generators 
essentially given by
\bea &&
[J_{ij}, J_{hk}] =2i  (\delta_{jk}J_{ih}+  \delta_{ih}J_{jk}  -\delta_{ik}J_{jh} - \delta_{jh}J_{ik}) \;,
\sea
[J_{ij}, X_k] = -2i(\delta_{jk} X_i - \delta_{ik} X_j) \;,
\qquad
[J_{ij}, P_k] = -2i(\delta_{jk} P_i - \delta_{ik} P_j) \;,
\sea
[J_{ij}, K_k] = -2i(\delta_{jk} K_i - \delta_{ik} K_j) \;,
\qquad
[K_i, K_j] = 
 0 \;,
\nonumber \\ &&
[K_i, H] = -2iP_i \;,
\quad\;
[K_i, P_j] = 
 0 \;,
\quad\
[X_i,P_j]= 2i\delta_{ij} I' \;,
\nonumber \\ &&
[T , H]= -2i I' \;,
\quad\quad
[K_i, T ] =
 0 \;,
\quad\;\;
[K_i, X_j]=  -2i\delta_{ij}  T \;.
\eea
Note that the full result for the other commutators beyond the $J_{ij}$
and $K_i$ set, originated from $SO(1,3)$, is essentially fixed by the
requirement of having the Galilean $K_i$-$H$ and the Heisenberg $X$-$P$
commutators. However, for the purpose here, the explicit contraction is 
to be implemented a bit differently. It is taken as the $c \to \infty$ 
limit of $K_i= \frac{1}{c} J_{i{\ssc 0}}$,  $P_i = \frac{1}{c} E_i$,  
$X_i=\frac{1}{c} Y_i$,$T =\frac{-1}{c^2} Y_{\ssc 0}$, $I'=\frac{1}{c^2} I$, 
with the renaming $H\equiv -E_{\ssc 0}$.  In the contraction, 
$K_i$ as generators for the Galilean boosts are the basic starting 
point and we would like to be able to trace physics, including the 
relative physical dimensions of quantities, by considering the 
speed of light $c$ as having a physical dimension. Introducing 
$X_i=\frac{1}{c} Y_i$ is to keep the same physical dimensions 
for $X_i$ and $P_i$. However, the essence of the contraction 
scheme as a formulation to retrieve an approximate physical 
theory from a more exact one is really to implement the 
contraction at a representation level.

To implement the contraction on  ${V}^{\ssc\! L}_{\!\ssc \varsigma}$, 
or the matching ${V}_{\!\ssc \varsigma}$ as a representation 
of the original $H\!(1,3)$, it is important to note that the original 
central charge generator $I$ represented by $\varsigma \hat{I}$
in ${V}_{\!\ssc \varsigma}$ would give the representation of the 
contracted $I'$, which remains central, as $\frac{\varsigma}{c^2} \hat{I}$. 
For a sensible result, one needs to consider $\varsigma=c^2  \chi$ 
with $\chi$ staying finite at the contraction limit, hence $I'$ 
represented by  $\chi \hat{I}$ (recall: $\hat{I}$ is the identity 
operator). Therefore,  ${V}_{\!\ssc \varsigma}$ contracts into  
${V}_{\!\ssc \chi}$. In another words, the ${V}_{\!\ssc \varsigma}$ 
representation of the original $H\!(1,3)$, and the full 
$H_{\!\ssc  R}(1,3)$, survives as the ${V}_{\!\ssc \chi}$ 
($\chi=\frac{\varsigma}{c^2}> 0$) representation of the 
$H\!(3)$ in the contracted $H_{\!\ssc G\!H}(3)$, as well 
as of the full group.

For the $c \to \infty$ limit of
${V}^{\ssc\! L}_{\!\ssc \chi}\!(p\vsi^\mu,x\vsi^\mu)$,  
we have to consider first
\[  
\hat{P}^{\ssc\! L}_{\!\ssc \chi i} = \frac{1}{c} \hat{E}^{\ssc\! L}_{\varsigma i} \;, 
\quad 
\hat{X}^{\ssc\! L}_{\!\ssc \chi i} = \frac{1}{c} \hat{Y}^{\ssc\! L}_{\varsigma i}\;, 
 \quad
\hat{H}^{\ssc\! L}_{\!\ssc \chi} =  -\hat{E}^{\ssc\! L}_{\varsigma\ssc 0}\;,
\quad
\hat{T}^{\ssc\! L}_{\!\ssc \chi} = -\frac{1}{c^2} \hat{Y}^{\ssc\! L}_{\varsigma \ssc 0} \;,
\] 
and take that to obtain
\bea && \label{resc}
\hat{X}^{\ssc\! L}_{\!\ssc (\!\chi\!) i} = \frac{1}{\sqrt{\chi}} \hat{X}^{\ssc\! L}_{\!\ssc \chi i}
= \hat{X}^{\ssc\! L}_{(\!\varsigma\!)i} \;,
\qquad
\hat{P}^{\ssc\! L}_{\!\ssc (\!\chi\!) i} = \frac{1}{\sqrt{\chi}} \hat{P}^{\ssc\! L}_{\!\ssc \chi i}
= \hat{P}^{\ssc\! L}_{(\!\varsigma\!)i} \;,
\sea
\hat{T}^{\ssc\! L}_{\!\ssc (\!\chi\!)} = \frac{1}{\sqrt{\chi}} \hat{T}^{\ssc\! L}_{\!\ssc \chi}
= -\frac{1}{c} \hat{X}^{\ssc\! L}_{(\!\varsigma\!)\ssc 0} \;,
\qquad
\hat{H}^{\ssc\! L}_{\!\ssc (\!\chi\!)} = \frac{1}{\sqrt{\chi}} \hat{H}^{\ssc\! L}_{\!\ssc \chi}
=- c \hat{P}^{\ssc\! L}_{(\!\varsigma\!)\ssc 0} \;,
\eea
(with $\varsigma = c^2 \chi$). The above are 
the basic set of operators acting on the functional space of 
 $\phi (p_{\!\ssc (\!\varsigma\!)},x_{\ssc (\!\varsigma\!)})$, 
with the variables properly rescaled to a new set of
variables to match with the operators. There is also the exactly 
corresponding set of operators, $\hat{X}_{{\!\ssc (\!\chi\!)}i}$,
$\hat{P}_{{\!\ssc (\!\chi\!)}i}$, $\hat{T}_{{\!\ssc (\!\chi\!)}}$,
and $\hat{H}_{{\!\ssc (\!\chi\!)}}$, and ${V}_{\!\ssc \chi}$
on the abstract Hilbert space which are helpful for tracing
 the proper description. The proper labels for the states  
$\left|p\vsi^\mu,x\vsi^\mu\rra$
at the contraction limit should be 
$\left| p^i_{\!\ssc (\!\chi\!)}, {e}_{\!\ssc (\!\chi\!)}, 
      x^i_{\!\ssc (\!\chi\!)}, {t}_{\!\ssc (\!\chi\!)} \rra$,
satisfying
\begin{align}
    2 x _{{\!\ssc (\!\chi\!)}i} =& 
 \prescript{}{_\eta}{\!\lla  p^i_{\!\ssc (\!\chi\!)}, {e}_{\!\ssc (\!\chi\!)},   x^i_{\!\ssc (\!\chi\!)}, {t}_{\!\ssc (\!\chi\!)}    
      \left| \hat{X}_{{\!\ssc (\!\chi\!)}i} \right| 
    p^i_{\!\ssc (\!\chi\!)}, {e}_{\!\ssc (\!\chi\!)},   x^i_{\!\ssc (\!\chi\!)}, {t}_{\!\ssc (\!\chi\!)}\rra }\;,
    \nonumber\\[3pt]
2 p _{{\!\ssc (\!\chi\!)}i} =& 
 \prescript{}{_\eta}{\!\lla  p^i_{\!\ssc (\!\chi\!)}, {e}_{\!\ssc (\!\chi\!)},   x^i_{\!\ssc (\!\chi\!)}, {t}_{\!\ssc (\!\chi\!)} 
         \left| \hat{P}_{{\!\ssc (\!\chi\!)}i} \right| 
   p^i_{\!\ssc (\!\chi\!)}, {e}_{\!\ssc (\!\chi\!)},   x^i_{\!\ssc (\!\chi\!)}, {t}_{\!\ssc (\!\chi\!)} \rra} \;,
 \nonumber\\[3pt]
2{t} _{\!\ssc (\!\chi\!)} =& 
  \prescript{}{_\eta}{\!\lla  p^i_{\!\ssc (\!\chi\!)}, {e}_{\!\ssc (\!\chi\!)},   x^i_{\!\ssc (\!\chi\!)}, {t}_{\!\ssc (\!\chi\!)}    
      \left| \hat{T}_{\!\ssc (\!\chi\!)} \right| 
   p^i_{\!\ssc (\!\chi\!)}, {e}_{\!\ssc (\!\chi\!)},   x^i_{\!\ssc (\!\chi\!)}, {t}_{\!\ssc (\!\chi\!)}\rra} \;,
 \nonumber\\[3pt]
2 {e} _{\!\ssc (\!\chi\!)} =& 
  \prescript{}{_\eta}{\!\lla p^i_{\!\ssc (\!\chi\!)}, {e}_{\!\ssc (\!\chi\!)},   x^i_{\!\ssc (\!\chi\!)}, {t}_{\!\ssc (\!\chi\!)}       
   \left| \hat{H}_{\!\ssc (\!\chi\!)} \right| 
   p^i_{\!\ssc (\!\chi\!)}, {e}_{\!\ssc (\!\chi\!)},  x^i_{\!\ssc (\!\chi\!)}, {t}_{\!\ssc (\!\chi\!)}\rra} \;,
\end{align}
and hence giving naively
\[
\phi(p\vsi^\mu,x\vsi^\mu)  
\quad \longrightarrow  \quad
\phi( p_{\!\ssc (\!\chi\!)}^i,{e}_{\!\ssc (\!\chi\!)},
   x_{\ssc (\!\chi\!)}^i,{t}_{\ssc (\!\chi\!)}\!)
\]
with 
\begin{align}\label{newpar}
    x _{\!\ssc (\!\chi\!)i} &= x _{\!\ssc (\!\chi\!)}^i = x_{\ssc (\!\varsigma\!)}^i \;,
&p _{\!\ssc (\!\chi\!)i} = p _{\!\ssc (\!\chi\!)}^i = p_{\ssc (\!\varsigma\!)}^i \;,
\nonumber\\
{t} _{\!\ssc (\!\chi\!)} &=  \frac{1}{c}\, x_{\ssc (\!\varsigma\!)}^{\ssc 0} \;,
&{e}_{\!\ssc (\!\chi\!)} = c\, p_{\ssc (\!\varsigma\!)}^{\ssc 0} \;. 
\end{align} 
 We have then, at least formally,
\begin{align}
    \hat{X}^{\ssc\! L}_{{\!\ssc (\!\chi\!)}i} =\:& x_{{\!\ssc (\!\chi\!)}i} + i \partial_{p^i_{\!\ssc (\!\chi\!)}} \;, &
&\hat{P}^{\ssc\! L}_{{\!\ssc (\!\chi\!)}i} =\: p_{{\!\ssc (\!\chi\!)}i} - i \partial_{x^i_{\!\ssc (\!\chi\!)}} \;,
\nonumber\\
\hat{T}^{\ssc\! L}_{\!\ssc (\!\chi\!)} =\:&  {t}_{\!\ssc (\!\chi\!)} - i \partial_{{e}_{\!\ssc (\!\chi\!)}} \;, &
&\hat{H}^{\ssc\! L}_{\!\ssc (\!\chi\!)}  =\: {e}_{\!\ssc (\!\chi\!)} + i  \partial_{{t}_{\!\ssc (\!\chi\!)}} \;.
\label{XPTH}
\end{align}
The crucial quantities controlling the nature of the representation are the overlaps
\[  
 \prescript{}{_\eta}{\!\lla  p^i_{\!\ssc (\!\chi\!)B}, {e}_{\!\ssc (\!\chi\!)B},   x^i_{\!\ssc (\!\chi\!)B}, {t}_{\!\ssc (\!\chi\!)B}    |
   p^i_{\!\ssc (\!\chi\!)A}, {e}_{\!\ssc (\!\chi\!)A},   x^i_{\!\ssc (\!\chi\!)A}, {t}_{\!\ssc (\!\chi\!)A}\rra }\;.
\]
From the original
$ \prescript{}{_\eta}{\!\lla  p^\mu_{\!\ssc (\!\chi\!)B},x^\mu_{\ssc (\!\chi\!)B}|    
    p^\mu_{\!\ssc (\!\chi\!)A}, x^\mu_{\ssc (\!\chi\!)A}\rra}$, 
given in Eq.(\ref{molap}), we have it as
\[
e^{i \left(  {e}\shib {t}\shia -t\shib {e}\shia
    +\delta_{ij}  x\shib^{i }p^j\shia-\delta_{ij} p^{i}\shib x\shia^j \right) }
   e^{-\frac{1}{2}\left[\left(x^{i}\shib-x^{i}\shia \right)^2 
 - c^2\left( {t}\shib-{t}\shia \right)^2 +\left(p^{i}\shib-p^{i}\shia \right)^2 
    -\frac{1}{c^2}\left({e}\shib-{e}\shia  \right)^2\right]}
\]
to be taken at the  $c \to \infty$ limit. It holds
$e^{\frac{1}{2c^2}\left({e}\shib-{e}\shia  \right)^2} \rightarrow 1$,
but the $e^{\frac{c^2}{2}\left({t}\shib-{t}\shia  \right)^2}$ 
factor diverges in the limit, except for ${t}\shib={t}\shia$, which 
indicates that we should consider only the latter case. The magnitude 
of the overlap being independent of ${e}\shib$ and ${e}\shia$ is still 
puzzling. The answer to that comes from a more careful thinking about 
the nature of the variables ${e}\shi$. Unlike ${t}\shi = \frac{x^0\vsi}{c}$, 
which is to be taken to be finite as in the general spirit of 
symmetry contraction, ${e}\shi = c \,p^{\ssc 0}$ is of quite 
different nature. The Lie algebra contraction to begin with 
only has a relabeling $H= - E_{\ssc 0}$  involving no $c$.
One may wonder if the $c$ in  
$\hat{H}^{\ssc\! L}_{\!\ssc (\!\chi\!)}
      =- c \hat{P}^{\ssc\! L}_{(\!\varsigma\!)\ssc 0}$
should be taken as giving a diverging energy observable 
$\hat{H}^{\ssc\! L}_{\!\ssc (\!\chi\!)}$ for any
finite $\hat{P}^{\ssc\! L}_{(\!\varsigma\!)\ssc 0}$.
Furthermore, for an Einstein particle of the rest mass $m$, 
{{\em i.e} the particle in Einstein's theory of
special relativity,}
\[
e = m c^2 + \frac{p_ip^i}{2m} + \cdots 
\]
where the neglected terms involve negative powers of $c^2$.
At the $c \to \infty$ limit, it is indeed diverging. Even
$p^{\ssc 0}$ is diverging. That is the result of the rest
mass as an energy. Hence, it sure suggests that we 
should take our variable ${e}\shi$ as infinite, and the 
`non-relativistic' energy we are interested in is the 
kinetic energy $\frac{p_ip^i}{2m}$ given by the limit of
$e - m c^2$. Taking that feature into our consideration, 
the Hilbert space of interest under the contraction is 
really only the space spanned by the $H(3)$ coherent states 
$\left| p^i_{\!\ssc (\!\chi\!)},  x^i_{\!\ssc (\!\chi\!)} \rra$
for a fixed time $t\shi$ and a formally infinite $e\shi$.
To be exact, we should be implementing that logic 
from an Einstein particle to our quantum observables 
$\hat{H}_{\ssc (\!\chi\!)}$,  $\hat{P}_{\ssc (\!\chi\!)}^{\ssc 0}$,
and $\hat{P}_{\ssc (\!\chi\!)i}$ or their expectation
values, but the conclusion is the same.  
The coherent state wavefunction
  $\phi_{\!\ssc A} (p^\mu\vsi,x^\mu\vsi)$ is equal to
$ \prescript{}{_\eta}{\!\lla  p^\mu_{\!\ssc (\!\varsigma\!)},x^\mu_{\ssc (\!\varsigma\!)}|
          p^\mu_{\!\ssc (\!\varsigma\!)A}, x^\mu_{\ssc (\!\varsigma\!)A} \rra}$,
hence at the contraction limit there is no more 
dependence on $t\shi$ and  $e\shi$, reducing it  
essentially to just $\phi_{\!\ssc A} (p^i\shi,x^i\shi)$ . 
The operator $\hat{T}^{\ssc\! L}_{\!\ssc (\!\chi\!)}$ 
acts on the space of wavefunctions only as 
a multiplication by $t\shi$ and is just like classical, 
while $\hat{H}^{\ssc\! L}_{\!\ssc (\!\chi\!)}$ is not
physically relevant. Note that the full contracted 
representation is then simply unitary. The part of the
inner product $ \prescript{}{_\eta}{\!\lla  \cdot | \cdot \rra}$ 
independent of $p^{\ssc 0}\vsi$ and $x^{\ssc 0}\vsi$, 
hence $t\shi$ and $e\shi$, is exactly the usual one,
{\em i.e.} $\hat\eta$ essentially reduces to identity
under the contraction. The space of wavefunctions 
spanned by $\phi_{\!\ssc A} (p^i\shi,x^i\shi)$ is
a Hilbert space.

\section{Group Theoretically Based WWGM Framework with
Wavefunctions in Coherent State Basis} \label{sec5}
The above analysis gives a successful picture of
the phase space of the  $H_{\!\ssc R}(1,3)$ theory, 
giving in the Galilean limit the phase space of the
$H_{\!\ssc R}(3)$ theory at each fixed `time' value.
The infinite dimensional manifolds give, at the proper
relativity symmetry contraction limit, the familiar 
finite dimensional classical models as approximation. 
The explicit results of the classical limit for the 
present case is presented in the section below. The 
merit of our group theoretical approach is that it 
gives a full dynamical theory associated with the 
corresponding spacetime/phase space model for 
each relativity symmetry, mutually connected through 
the contraction/deformation pattern. The dynamical
theory is naturally a Hamiltonian theory from the 
symmetry of the phase space as symplectic geometry.
The dynamics is better described on the algebra of 
observables as essentially the matching representation 
of the group C$^*$-algebra  \cite{066,070,082}.  
Moreover, all those fit in well with the idea of the
position and momentum operators as noncommutative 
coordinates of the phase space \cite{082,078,081}. 

\subsection{The Algebra of Observables, Symmetries, and Dynamics}
The algebra of observables is depicted essentially as 
the one from a WWGM formalism, as functions and 
distributions of the position and momentum operators 
$\hat{X}_\mu$ and $\hat{P}_\mu$. The basic dynamical 
variables of our representation on the space of 
wavefunctions $\phi(p^\mu\vsi,x^\mu\vsi)$ are 
$\hat{X}^{\!\ssc L}= x+ i\partial_{p}= x\star$ and 
$\hat{P}^{\!\ssc L} = p - i\partial_{x}= p\star$,
where we have dropped the $\mu$ indices and the 
subscript $\vsi$.  We may also write a general function 
of $(p^\mu\vsi,x^\mu\vsi)$ as simply $\za(p,x)$, 
and the $\star$ is as in the Moyal star product 
\bea
\za \star \zb (p,x) = \za(p,x) e^{-i (\cev{\partial}_p \vec{\partial}_x-\cev{\partial}_x \vec{\partial}_p) } \zb(p,x) \;,
\eea
with  $\za(p,x)\star=\za(p\star,x\star)$. Under such 
notation, the story looks quite the same as the case 
for $H_{\!\ssc R}(3)$ with only  $\hat{X}_i^{\!\ssc L}$, 
and  $\hat{P}_i^{\!\ssc L}$ as $x_i\star$ and $p_i\star$, 
given in details in Ref.\cite{070}.  Hence, we present 
here only a summary of the results, leaving the readers 
to consult the latter paper and references therein. 

Let us take a little detour first to clarify our theoretical 
perspective. What we have is rather like the WWGM 
put up-side-down \cite{070}. We start with the quantum 
theory as an irreducible representation of a (quantum) 
relativity symmetry, including the Heisenberg-Weyl 
symmetry. With the wavefunction in the coherent 
state basis as the natural reduction of the representation 
of the group algebra, the corresponding representation 
of the latter properly extended serves as the algebra 
of observables. The latter can be seen as a collection 
of functions and tempered distribution of  the position 
and momentum operators represented as differential 
operators by $x\star$ and $p\star$. The real variables 
$x$ and $p$ are not quite the coordinates of the 
classical phase space. Only their rescaled counterparts 
under the contraction of the symmetry to the classical 
relativity symmetry are. Contrary to a deformation 
quantization, a contraction is a de-quantization 
procedure. From the algebraic point of view, the 
deformation of an observable algebra as in WWGM 
is really a result of a deformation of the classical 
relativity symmetry to the quantum one, pushed 
onto the group C$^*$-algebra of the symmetry. The 
contraction is exactly the inverse of the deformation
 \cite{060}, at a Lie algebra level and beyond.

In the usual unitary quantum mechanics, on the Hilbert 
space ${\mathcal{K}}$ of wavefunctions $\phi(p,x)$,  
symmetries are represented in a form of unitary and 
antiunitary operators, factored by its closed center 
of phase transformations. On the set ${\mathcal{P}}$ 
of pure state density operators $\rho_\phi(p,x)\star$, 
corresponding to the abstract projection operator
$\hat\rho_\phi = \left|\phi\rra\!\lla \phi \right|$ for 
normalized $\left|\phi\rra$, the automorphism group 
$Aut({\mathcal{P}})$  is characterized by the subgroup 
of the group of real unitary transformations 
${\mathcal{O}}(\tilde{\mathcal{K}}_{\!\ssc R})$
compatible with the star product,  
$\tilde{\mathcal{K}}_{\!\ssc R}$ being the real 
span of  all $\rho_\phi(p,x)\star$, the complex 
extension of which is the Hilbert space of
Hilbert-Schmidt operators, as in the Tomita 
representation. We write the unitary 
transformations in the form
\[
 \tilde{U}_\star\za\star=\mu(\za)\star=U_\star\! \star \za  \star  \bar{U}_\star\!\star \;,
\]
with $\mu \in Aut({\mathcal{P}})$, where
${U}_\star\!\star \equiv {U}_\star(p,x)\star$
is a unitary operator on ${\mathcal{K}}$, generated 
by the Hermitian operator in the form of a real function 
$G_{\!s}(p\star,x\star)$, and $\bar{U}_\star\!\star$ 
is its inverse obtained by the complex conjugation and
$ \tilde{U}_\star \in {\mathcal{O}}(\tilde{\mathcal{K}}_{\!\ssc R})$. 
We refer to the ${U}_\star\!\star$ as star-unitary, in 
particular whenever necessary to highlight it being 
a function of the  $p\star$ and $x\star$ operators.

The above, illustrated for the case of  $H_{\!\ssc R}(3)$
formulation of standard quantum mechanics in Ref. \cite{070},
can be applied to our $H_{\!\ssc R}(1,3)$ case with a slight 
modification. We need to use the invariant inner product with 
$\hat\rho_\phi = \left|\phi\rra\! \prescript{}{_\eta}{\!\lla \phi \right|}$
for normalized $\left|\phi\rra$, and replace the Hermitian
 and unitary requirements by $\eta$-Hermitian and 
$\eta$-unitary ones.  Our relevant symmetry 
transformations are to be given by $\eta$-unitary 
operator $V_{\star(s)} \star$ generated by
$\eta$-Hermitian $G_{\!s}(p\star,x\star)$, which 
are real functions of  the basic $\eta$-Hermitian  
operators $(p\star,x\star)$,
{\em i.e.} $G_{\!s}(\hat{P}^{\!\ssc L}_{\!\ssc \mu},\hat{X}^{\!\ssc L}_{\!\ssc \mu})
   = \overline{G}_{\!s}(\hat{P}^{\!\ssc L}_{\!\ssc \mu},\hat{X}^{\!\ssc L}_{\!\ssc \mu})$,
and we use the $\bar\za$ to denote the `complex 
conjugate' of  $\za$ as a function which correspond
 to $\bar\za\star$ as the $\eta$-Hermitian conjugate 
of $\za\star$ as an operators as an element of the
observable algebra. The conjugation is the involution 
of the latter as a $^*$-algebra.
$\overline{V}_{\star(s)}\star$ of a $\eta$-unitary 
${V}_{\star(s)}\star$ is to be interpreted in the same
 manner. The feature of $\overline{V}_{\star(s)}\star$ 
to be the inverse of ${V}_{\star(s)}\star$ is exactly  
$\eta$-unitarity. Again, $\eta$-Hermiticity is the
Hermiticity so long as the algebraic analysis is
concerned. Though $\eta$-unitarity here is really
pseudo-unitarity, only the inner product preserving
nature of it is relevant here and it is as good as 
unitarity. And of course Krein spaces are to be
allowed in the place of the Hilbert spaces.

Generators of our relativity symmetry $H_{\!\ssc R}(1,3)$ 
are to be represented as a subgroup of $Aut({\mathcal{P}})$ 
of the observable algebra. All $H_{\!\ssc R}(1,3)$ 
generators are $\eta$-Hermitian, hence each is given 
by a real $G_{\!s}$, generating (star-)$\eta$-unitary
${V}_{\star(s)}\!\star=e^{\frac{-i s}{2}{G}_{\!s}\star}$ 
as one-parameter groups of symmetry transformations.
Note that the factor $2$ is really $\hbar$. We have 
$\tilde{V}_{\star(s)}= e^{\frac{-i s}{2}\tilde{G}_{\!s}}$, 
\bea\label{vg}
 \tilde{V}_\star\za\star=\mu(\za)\star=V_\star\! \star \za  \star  \overline{V}_\star\!\star 
\eea
with 
\bea \label{tildeG}
\tilde{G}_{\!s} \rho 
    ={G}_{\!s}\!\star \rho -\rho \star\!{G}_{\!s} 
  = 2i \{{G}_{\!s},  \rho\}_\star\;,
\eea
where $\rho(p,x)\in  \tilde{\mathcal{K}}$ and 
$\{\cdot, \cdot\}_\star$  is the Moyal bracket. 
Hence, with $\rho(s)=\tilde{V}_{\star(s) }\rho(s=0)$, 
\bea \label{leq}
\frac{d}{ds}\rho(s)= \{{G}_{\!s},  \rho(s)\}_\star \;.
\eea
The equation  is the Liouville equation of motion for a 
mixed state $\rho$ in $\tilde{\mathcal{D}}$, the self-dual 
cone of $\tilde{\mathcal{K}}$. The class of operators on 
$\tilde{\mathcal{K}}$ representing symmetry generators 
are important, especially for tracing the symmetries to
 the classical limit where all ${G}_{\!s}\star$ reduce 
essentially to the commutative ${G}_{\!s}$, as multiplicative 
operators on the functional space of classical observables.
We can write $\tilde{G}_{\!s}=\hat{G}_{\!s}^{\ssc L}-\hat{G}_{\!s}^{\ssc {R}}$, 
where $\hat{G}_{\!s}^{\ssc L}\equiv {G}_{\!s}(p,x)\star
      ={G}_{\!s}(\hat{P}^{\!\ssc L},\hat{X}^{\!\ssc L})$ 
is a left action and $\hat{G}_{\!s}^{\ssc {R}}$
 is the corresponding right action defined by  
$\hat{G}_{\!s}^{\ssc  {R}}\za\equiv \za\star\!{G}_{\!s}(p,x) 
    ={G}_{\!s}(\hat{P}^{\!\ssc  {R}},\hat{X}^{\!\ssc  {R}})\za$.
Analogously to $\hat{X}^{\!\ssc  L}$ and $\hat{P}^{\!\ssc  L}$ 
coming from the left-invariant vector fields of the Heisenberg-Weyl 
group, there are those from the right-invariant ones given by
\bea
\hat{X}^{\!\ssc  {R}}= x -  i \partial_{p}\;,
\qquad
\hat{P}^{\!\ssc  {R}} = p +  i \partial_{x}\;.
\label{R}
\eea

From Eq.(\ref{v-shift}) we see that
\bea \label{phitrans}
V_{\star(-x'^\mu)}\!\star \phi (p^\mu,x^\mu)& =& e^{\frac{-i x'^\mu}{2} (-p_\mu\star)} \phi (p^\mu,x^\mu)
  = \phi \!\left(p^\mu,x^\mu+\frac{x'^\mu}{2} \right) e^{\frac{ix'_\mu p^\mu}{2}}\;,
\nonumber \\
V_{\star(p'^\mu)}\!\star \phi (p^\mu,x^\mu)& =& e^{\frac{-i p'^\mu}{2}(x_\mu\star)} \phi (p^\mu,x^\mu) 
  =  \phi \!\left(p^\mu+\frac{p'^\mu}{2},x^\mu \right) e^{\frac{-ip'_\mu x^\mu}{2}}\;.
\eea
In the above,  for the wavefunctions, we show only the 
involved pair of variables in each case, and there is always 
no summation over indices. The other variables are simply
not affected by the transformations. In terms of the 
parameters  $x^\mu$ and $p^\mu$, we have
\bea
G_{\!-x^\mu}\star &=& p_\mu\star\;, 
\qquad\qquad 
\tilde{G}_{\!-x^\mu} =-2i \partial_{x^\mu} \;, 
\nonumber \\
G_{\!p^\mu}\star &=& x_\mu\star \;, 
\qquad\qquad   
\tilde{G}_{\!p^\mu} = 2i \partial_{p^\mu} \;,
\eea
all in the same form as in the $H_{\!\ssc R}(3)$ case.
The factors of $2$ in the translations $V_{\star(x)}\!\star$ 
and $V_{\star(p)}\!\star$, though somewhat suspicious 
at the first sight, are related to the fact that the arguments 
of the wavefunction correspond to half of the expectation 
values, due to our coherent state labeling. Thus, $x_\mu\star$ 
and $p_\mu\star$ generate translations of the expectation 
values, which is certainly the right feature to have.
For the Lorentz transformations, we have 
$G_{\!\omega^{\mu\nu}}=(x_\mu p_\nu-x_\nu p_\mu)$,
\bea
G_{\!\omega^{\mu\nu}}\star &=& (x_\mu p_\nu -i x_\mu \partial_{x^\nu} + i p_\nu \partial_{p^\mu}
    + \partial_{x^\nu} \partial_{p^\mu}) - (\mu \leftrightarrow \nu)\;,
\nonumber \\
\tilde{G}_{\!\omega^{\mu\nu}}  &=& -2i (x_\mu \partial_{x^\nu}- p_\nu \partial_{p^\mu} )- (\mu \leftrightarrow \nu)\;.
\eea
with the explicit action (no summation over the indices)
\begin{equation}
V_{\star(\omega^{\mu\nu})}\!\star \phi (p,x)
 = e^{\frac{-i\omega^{\mu\nu}}{2} (G_{\!\omega^{\mu\nu}}\star)} \phi (p,x)
    = \phi \!\left(\! e^{\frac{i\omega^{\mu\nu}}{2}  \widehat{G}_{\!\omega^{\mu\nu}} }[p,x] \!\right) ,
\end{equation}
where $\widehat{G}_{\!\omega^{\mu\nu}}$ are the  
infinitesimal $SO(1,3)$ transformation operators 
corresponding to the coset space action to be obtained
from Eq.(\ref{group}). 

All the  $G_{\!-x^\mu},\,G_{\!p^\mu}$ and $G_{\!\omega^{\mu\nu}}$ 
(and $G_{\!\theta}=1$) make the full set of operators 
for the generators  $\hat{G}_{\!s}^{\ssc L}=G_{\!s}\star$
of the $H_{\!\ssc R}(1,3)$ group representing the
symmetry on ${\mathcal{K}}$, and constitute a 
Lie algebra within the algebra of physical observables.
$\hat{G}_{\!s}^{\ssc R}$ set does the same as a right 
action, and $\hat{G}_{\!s}^{\ssc L}$ always commute 
with  $\hat{G}_{\!s'}^{\ssc R}$ since, in general,
$[ \hat\za^{\!\ssc L}, \hat\zc^{\!\ssc {R}}]=0$. 
These fourteen $G_{s}$ as multiplicative operators, 
of course, all commute among themselves. The  
commutators for $\tilde{G}_{\!s}$ are same as those 
for $\hat{G}_{\!s}^{\ssc L}$, with however the 
vanishing $\tilde{G}_{\!\theta}$  giving a vanishing 
$[\tilde{G}_{\!p^\mu}, \tilde{G}_{\!-x^\nu}]$.  
For any function $\za(p^\mu,x^\mu)$, there are four 
associated operators on $\tilde{\mathcal{K}}$. Those 
are $\za, \hat{\za}^{\!\ssc L},\hat{\za}^{\!\ssc {R}}$ 
and $\tilde{\za}$, but only two of them are linearly 
independent. For our relativity symmetry operators, 
the independent set $\{G_{\!-x^\mu},G_{\!p^\mu},G_{\!\omega^{\mu\nu}},
   \tilde{G}_{\!-x^\mu},\tilde{G}_{\!p^\mu},\tilde{G}_{\omega^{\mu\nu}}\}$
has the only non-vanishing commutators among 
them given by (we also have ${G}_{\!\theta}=1$, 
the identity, and  $\tilde{G}_{\!\theta}=0$)
\bea &&
[{G}_{\!\omega^{\mu\nu}}, \tilde{G}_{\!\omega^{\za \zb}}]
   = 2i( \eta_{\nu \zb} G_{\!\omega^{\mu \za }} -  \eta_{\nu \za } G_{\!\omega^{\mu \zb}}
                   +  \eta_{\mu \za } G_{\!\omega^{\nu  \zb}}-   \eta_{\mu \zb} G_{\!\omega^{\nu  \za }} ) \;,
\nonumber \\ &&
[{G}_{\!\omega^{\mu\nu}}, \tilde{G}_{\!-x^\za}] 
  =  -2i( \eta_{\nu\za} G_{\!-x^\mu} - \eta_{\mu \za} G_{\!-x^\nu} ) \;,
\nonumber \\ &&
[{G}_{\!\omega^{\mu \nu}}, \tilde{G}_{\!p^\za}]
  = -2i( \eta_{\nu \za} G_{\!p^\mu} - \eta_{\mu  \za} G_{\!p^\nu} ) \;,
\nonumber \\ &&
[\tilde{G}_{\!\omega^{\mu\nu}}, {G}_{\!-x^\za}]
   = -2i( \eta_{\nu\za} G_{\!-x^\mu} - \eta_{\mu \za} G_{\!-x^\nu} )  \;,
\nonumber \\ &&
[\tilde{G}_{\!\omega^{\mu\nu}}, {G}_{\!p^\za}]
   = -2i( \eta_{\nu \za} G_{\!p^{\mu}} - \eta_{\mu\za} G_{\!p^\nu} ) \;,
\nonumber \\ &&
[{G}_{\!p^\mu}, \tilde{G}_{\!-x^\nu}] =- [{G}_{\!-x^\mu}, \tilde{G}_{\!p^\nu}]
   = 2i \eta_{\mu\nu} \;,
\nonumber \\ &&
[{G}_{\!p^\mu}, \tilde{G}_{\!p^\nu}] =
[{G}_{\!-x^\mu}, \tilde{G}_{\!-x^\nu}] = 0\;.
\label{GtG}
\eea

Quantum dynamics is completely symplectic, whether
described in the Schr\"odinger picture in terms of
real/complex coordinates of the (projective) Hilbert
space or the Heisenberg picture as a description in
terms of the noncommutative coordinates \cite{078}. 
The explicit dynamical equation of motion is to be 
seen as the transformations generated by a physical 
Hamiltonian characterized by an evolution
parameter. In the $H_{\!\ssc R}(3)$ case of the 
usual (`non-relativistic') quantum mechanics, it is  
${G}_{\!t }= \frac{p_ip^i}{2m}$. 
For our  $H_{\!\ssc R}(1,3)$ case, we consider 
${G}_{\!\tau}=\frac{p_\mu p^\mu}{2m}$ with the 
parameter $\tau$ being the Einstein proper time, 
which is expected to give the standard covariant 
description of Einstein particle dynamics, as we  
see explicitly below.

For some $s$-dependent operator $\za(p^\mu(s),x^\mu(s)) \star$ 
and a general Hamiltonian $G_{\!s}$, Heisenberg 
equation of motion is given by
\begin{equation}
    \frac{d}{ds}\za \star =\frac{1}{2i} \left[\za \star,G_{\!s}\!\star \right] \;.
\end{equation}
The right-hand side of the equation is simply the 
Poisson bracket of  $\za(p\star,x\star)$ and 
$G_{\!s}(p\star,x\star)$, functions of the 
noncommutative canonical variables 
$p^\mu\star$ and $x^\mu\star$. The equation
can simply be written as 
\bea 
\frac{d}{ds}\za=\{\za,{G}_{\!s} \}_\star=\frac{-1}{2i}\tilde{G}_{\!s}\za \;,
\eea
and is exactly the differential version of the automorphism
flow given in Eq.(\ref{vg}), here with our $\eta$-unitary
symmetry flows ${V}_{\star(s)}\star= e^{\frac{-i s}{2}{G}_{\!s}\star}$
generated by a  $\eta$-Hermitian ${G}_{\!s}\star$. 
$\frac{-1}{2i}\tilde{G}_{\!s}$ is really a Hamiltonian vector 
field for a Hamiltonian function $G_{\!s}(p\star,x\star)$ \cite{078}.

Our physical Hamiltonian operator ${G}_{\!\tau}(p\star)$
is such a $\eta$-Hermitian ${G}_{\!s}\star$. The 
corresponding Heisenberg equation gives, in particular,
\bea &&
\frac{d}{d\tau}x_\mu\star = \frac{1}{2i} \frac{1}{2m}[x_\mu\star,p_\nu\star p^\nu\star]
=\frac{p_\mu\star}{m} =\frac{\partial G_{\!\tau}(p\star)}{\partial (p^\mu\star)} \;,
\sea
\frac{d}{d\tau}p_\mu\star = \frac{1}{2i}\frac{1}{2m}[p_\mu\star,p_\nu\star p^\nu\star]
= 0
=-\frac{\partial G_{\!\tau}(p\star)}{\partial (x^\mu\star)} \;,
\eea
which are exactly
\bea\label{heq}
\frac{d}{d\tau}  \hat{X}_\mu^{\!\ssc L} 
=\frac{\partial G_{\!\tau}(\hat{P}_\nu^{\!\ssc L}  )}{\partial \hat{P}^{{\!\ssc L}^\mu} } \;,
\qquad
\frac{d}{d\tau}  \hat{P}_\mu^{\!\ssc L} 
=-\frac{\partial G_{\!\tau}(\hat{P}_\nu^{\!\ssc L} )}{\partial \hat{X}^{{\!\ssc L}^\mu} } \;,
\eea
the standard form of Hamilton's equations 
of motion for the canonical $\eta$-Hermitian
operator coordinate pairs 
$\hat{X}_\mu^{\!\ssc L}$-$\hat{P}_\mu^{\!\ssc L}$.
As usual in a Hamiltonian formulation, the constant,
or $\tau$-independent, momentum $\hat{P}_\mu^{\!\ssc L}$ 
is obtained from the equations of motion as
velocity multiplied by the particle mass $m$. 
Here, $-m^2$ is just the constant value of 
$p_\nu p^\nu$ as $2mG_{\!\tau}$.

For the Schr\"odinger picture, as $\eta$-unitary
flows on $\mathcal{K}$, we have the equation
\begin{equation}
    \frac{d}{ds}\phi=\frac{1}{2i}G_{\!s}\star \phi \;,
\end{equation}
which  for $G_{\!\tau}\star$ gives the 
$\tau$-independent solution for $\phi$ in the 
exact form of the Klein-Gordon equation, provided 
that the $G_{\!\tau}\star$ eigenvalue is taken 
to be $-\frac{m}{2}$.  Explicitly, in terms of the 
basic variables $p^\mu$ and $x^\mu$, we have
\begin{equation}
    G_{\!\tau}\star \phi(p,x) 
=\frac{1}{2m} p_\mu\star p^\mu\star \phi(p,x) 
=\frac{1}{2m} \left(p^\mu p_\mu- \eta^{\mu\nu}\partial_{x^\mu}\partial_{x^\nu}
    -2ip^\mu\partial_{x^\mu}\right)\phi(p,x)
\;,
\end{equation}
giving the wavefunctions 
$\phi(p,x)=e^{i(2k_\mu-p_\mu) x^\mu}$ for eigenvalues 
$\frac{2k^\mu k_\mu}{m}$.  Eigenvalues of the momentum 
operators $p_\mu\star$ are $2k_\mu$, satisfying 
$(2k^\mu) (2k_\mu)=-m^2$.  The factor of 2 really 
corresponds to $\hbar$, as in the standard textbook 
expression. Finally, the $\tau$-dependence is then 
given by $\frac{d}{d\tau}\phi = -\frac{m}{4i} \phi$,
as expected.

\subsection{Lorentz to Galilean Contraction} \label{LtoG2}
Contraction to Galilean limit has been presented in 
Sec.~\ref{sec4} at the kinematical level. In this section, 
we present the corresponding contraction in the 
observable algebra given in the WWGM formalism. Recall 
that the original Krein space under the contraction becomes 
reducible into a sum of essentially identical irreducible 
components, each being spanned by the wavefunctions 
$\phi(p^i,x^i) \equiv \phi (p\shi^i, x\shi^i)$
for a particular value of `time' ${t}_{\!\ssc (\!\chi\!)}$.
A general operator 
$\za(\hat{X}_{\mu}^{\!\ssc L},\hat{P}_{\mu}^{\!\ssc L})$
should then be seen as 
$\za(\hat{X}_{i}^{\!\ssc L},\hat{P}_{i}^{\!\ssc L},
   \hat{T}^{\ssc\! L}_{\!\ssc (\!\chi\!)},
   \hat{H}^{\ssc\! L}_{\!\ssc (\!\chi\!)})$ with 
$\hat{X}_{i}^{\!\ssc L} \equiv \hat{X}^{\ssc\! L}_{\!\ssc (\!\chi\!)i}$
and $\hat{P}_{i}^{\!\ssc L} \equiv \hat{P}^{\ssc\! L}_{\!\ssc (\!\chi\!)i}$,
from results of Eq.(\ref{XPTH}). Hence, on $\phi(p^i,x^i)$ 
we have effectively Hermitian actions of operators
$\hat{X}_{i}^{\!\ssc L} = x_i + i \partial_{p^i}$,
$\hat{P}_{i}^{\!\ssc L} = p_i - i \partial_{x^i}$,
$\hat{T}^{\ssc\! L}_{\!\ssc (\!\chi\!)} \to {t}_{\!\ssc (\!\chi\!)}$,
and $\hat{H}^{\ssc\! L}_{\!\ssc (\!\chi\!)} \to {e}_{\!\ssc (\!\chi\!)}$,
with the last two reduced to a simple multiplication 
by the `variables'  ${t}_{\!\ssc (\!\chi\!)}$ and 
(formally infinite) ${e}_{\!\ssc (\!\chi\!)}$, 
respectively. All $\za(p^\mu\star,x^\mu\star)$ 
operators on $\phi(p^i,x^i)$ reduce to 
$\za(p^i\star,x^i\star, {t}\shi, {e}\shi)$, or rather 
simply to $\za(p^i\star,x^i\star)$ like in the basic 
quantum mechanics, a unitary representation theory of 
$H_{\!\ssc R}(3)$.  The $\star$ should now be seen as 
the one  involving only variables $p^i$ and $x^i$. 

The transformations generated by the Hermitian
$G_{\!-x^i}\star,G_{\!p^i}\star$ and $G_{\!\omega^{ij}}\star$ 
obviously do not change. They represent generators of 
the $H_{\!\ssc R}(3)$ subgroup of $H_{\!\ssc R}(1,3)$ 
to begin with. $\tilde{G}_{\!-x^i},\tilde{G}_{\!p^i}$ 
and $\tilde{G}_{\!\omega^{ij}}$ are also unchanged. 
$G_{\!-x^{\ssc 0}}\star$ and $G_{\!p^{\ssc 0}}\star$, 
representing $\hat{P}_{\ssc (\!\varsigma\!)0}^{\!\ssc L}$
and $\hat{X}_{\ssc (\!\varsigma\!)0}^{\!\ssc L}$,
are to be replaced under the contraction by
$\hat{H}^{\ssc\! L}_{\!\ssc (\!\chi\!)}$ and
$\hat{T}^{\ssc\! L}_{\!\ssc (\!\chi\!)}$, respectively, with
 $V_{\star(-x^{\ssc 0})} =e^{\frac{i x^{\ssc 0}}{2}G_{\!-x^{\ssc 0}}}$
and $V_{\star(p^{\ssc 0})} =e^{\frac{-i p^{\ssc 0}}{2}G_{\!p^{\ssc 0}}}$ 
re-expressed as $V_{\star(t)} =e^{-\frac{it}{2}G_{\!t}}$
and $V_{\star(e)} =e^{\frac{ie}{2}G_{\!-e}}$, 
where $G_{\!t}\star=\hat{H}^{\ssc\! L}_{\!\ssc (\!\chi\!)}$
and $G_{\!-e}\star=\hat{T}^{\ssc\! L}_{\!\ssc (\!\chi\!)}$. 
On the wavefunction $\phi(p^i,x^i)$, we have the infinite
$G_{\!t}\star={e}\shi$ and finite $G_{\!-e}\star={t}\shi$. 
We also have $\tilde{G}_{\!t}= 2i \partial_{{t}\shi}$ and
$\tilde{G}_{\!-e} = -2  i\partial_{{e}\shi}$. 
None of the four operators are of interest, so long 
as their action on the observable algebra for an 
irreducible representation $\phi(p,x)$ is concerned.

The other interesting ones to check are the Lorentz boosts 
under the contraction. The generator $J_{i{\ssc 0}}$ 
in the Lie algebra is replaced by the finite $K_i=
\frac{1}{c}J_{i{\ssc 0}}$ . The group elements
$e^{i{\omega^{i0}}J_{i{\ssc 0}}}$ are to be re-expressed as
$e^{i{\zb^i} K_i}$ with ${\zb^i}= c \, {\omega^{i0}}$. 
In the original representation, the $J_{i{\ssc 0}}$ action is given by 
$G_{\!\omega^{i0}}\star=
   \hat{X}_{\ssc (\!\varsigma\!)i}^{\!\ssc L}  \hat{P}_{\ssc (\!\varsigma\!)0}^{\!\ssc L}
    - \hat{X}_{\ssc (\!\varsigma\!)0}^{\!\ssc L}\hat{P}_{\ssc (\!\varsigma\!)i}^{\!\ssc L}  $, 
from which follows the action of $K_i$ as
\[
G_{\!\zb^i}\star =  \hat{X}_{i}^{\!\ssc L}\left(\frac{-1}{c^2}\hat{H}^{\ssc\! L}_{\!\ssc (\!\chi\!)} \right)-  \left(-\hat{T}^{\ssc\! L}_{\!\ssc (\!\chi\!)}\right)\hat{P}_{i}^{\!\ssc L}
\;   \to  \; {t}_{\!\ssc (\!\chi\!)} p_i\star 
   = {t}_{\!\ssc (\!\chi\!)} G_{\!-x^i}\star
\]
with $V_{\star(\zb^i)}= e^{\frac{-i \zb^i}{2} G_{\!\zb^i}}$
(no summation over $i$), a re-writing of
$V_{\star(\omega^{i0})}$ with the new finite parameter 
$\zb^i$. We have seen, in Eq.(\ref{phitrans}) explicitly,  
that $V_{\star(-x^i)}\star$ gives a translation in the 
variable $x^i$ of the wavefunction. $V_{\star(\zb^i)}\star$ 
is then a time variable ${t}_{\!\ssc (\!\chi\!)}$-dependent 
translation, a Galilean boost exactly as the Lie 
algebra contraction promised, and is now unitary.
Similarly, we have
\bea
\tilde{G}_{\!\zb^i} = \frac{1}{c}\tilde{G}_{\!\omega^{i0}}
 &=& -\frac{2i}{c^2} (x_i\partial_{t\shi} +{e\shi} \partial_{p^i})+2i ( -t\shi \partial_{x^i}-p_i \partial_{e\shi}) 
\sea
\rightarrow -2i (t\shi \partial_{x^i} +p_i \partial_{e\shi}) \;.
\eea
We keep the $\partial_{e\shi}$ since the $\tilde{G}_{\!\zb^i}$ 
may act on the mixed states. We have the newly relevant 
nonzero commutators involving a ${G}_{\!\zb^i}$, ${G}_{\!t}$, 
or $G_{\!-e}$, and a $\tilde{G}_{\!s}$ as well as those 
involving a $\tilde{G}_{\!\zb^i}$, $\tilde{G}_{\!t}$, 
or $\tilde{G}_{\!-e}$ and a ${G}_{\!s}$, all from the
generators of the Lie algebra, as
\bea &&
[{G}_{\!\zb^i}, \tilde{G}_{\!\omega^{\ssc jk}}]
=-2i\left(  \delta_{\ssc ij} G_{\!\zb^k}-\delta_{\ssc ik} G_{\!\zb^j}\right)\; ,
\sea  
 [{G}_{\!\omega^{\ssc ij}}, \tilde{G}_{\!\zb^k}]
=2i\left(  \delta_{\ssc ik} G_{\!\zb^j}-\delta_{\ssc jk} G_{\!\zb^i}\right)\;,  
\sea
[{G}_{\!\zb^i}, \tilde{G}_{\!t}]=[\tilde{G}_{\!\zb^i}, G_{\!t}]=-2iG_{\!-x^{\ssc i}} \; , 
\sea
[{G}_{\!\zb^i}, \tilde{G}_{\!p^{\ssc j}}]=[\tilde{G}_{\!\zb^i},{G}_{\!p^{\ssc j}}]=-2i\delta_{\ssc ij}G_{\!-e} \;,
\sea
[{G}_{\!-e}, \tilde{G}_{\!t}] 
=- [{G}_{\!t}, \tilde{G}_{\!-e}]=-2i \;.
\eea

Since on the Hilbert space of the contracted theory 
we have only $\phi(p^i,x^i)$ and the corresponding 
observable algebra as $\za(p_i\star,x_i\star)$, the 
loss of $p_{\ssc 0}\star$ and $x_{\ssc 0}\star$, the 
quantum observables of energy and time, means 
that the Heisenberg equation of motion, in the 
form of a differential equation in $\tau$, 
effectively corresponds to the part of 
$G_{\!\tau}\star$ involving only $p^i\star$. 
We have
\bea
\frac{d}{d\tau} \za\star = \frac{1}{2i} [\za\star, G_{\!\tau}\star]
= \frac{1}{2i} [\za\star, G_{\!t}\star] \;,
\eea
where $G_{\!t}= \frac{p_i p^i}{2m}$, giving 
the right time evolution in the `non-relativistic', 
or $H_{\!\ssc R}(3)$, quantum theory, as expected. 
At the $c \to \infty$ limit, the proper time is just 
the Newtonian time. One can also see that the 
quantum Poisson bracket $\frac{1}{2i}[\cdots,\cdots]$ 
does suggest that the now multiplicative operators 
$t\shi$ and $e\shi$, from the original $p_{\ssc 0}\star$ 
and $x_{\ssc 0}\star$, are to be dropped from the  
canonical coordinates of the noncommutative 
symplectic geometry, in line with the Hilbert 
space picture. A $G_{\!t}$ of the form 
$\frac{p_i p^i}{2m} + v(x^i)$, {\em i.e.} with a
nontrivial interaction potential of course cannot
be retrieved from a $G_{\!\tau}$ which does 
not allow that, so long as the Einstein theory is
concerned. If we allow a nontrivial $v(x^\mu)$
in $G_{\!\tau}$, however, everything 
works fine. For the latter $G_{\!\tau}$ to be
taken as a `relativistic' Hamiltonian, one would
have to allow violation of the Einstein 
relation of $p_\mu p^\mu= -m^2$.

\section{Contraction to Classical Theory in Brief} \label{sec6}

In this section we look at the corresponding classical theory 
at the Lorentz covariant level through the contraction 
along the line of the one performed in the `non-relativistic', 
$H_{\!\ssc R}(3)$, case presented in Ref.\cite{070}. Only 
a sketch will be presented where the mathematics is 
essentially the same with the latter. The contraction 
trivializing the commutators between the position and 
momentum operators is obtained by rescaling the 
generators as 
\bea
X_\mu^c = \frac{1}{k_x} X_\mu
\qquad \mbox{and} \qquad
\label{XPc}
P_\mu^c =  \frac{1}{k_p} P_\mu \;,
\eea
and taking the limit $k_x, k_p \to \infty$. The only 
important difference between $k_x$ and $k_p$ 
parameters is their physical dimensions, giving the 
$X_\mu^c$ and $P_\mu^c$ observables with their 
different classical units. For the corresponding 
operators we have
\bea
\hat{X}^{c\ssc L} &=&    x^c +   i \frac{1}{k_x k_p} \partial_{p^c} \longrightarrow x^c\;,
\nonumber \\
\hat{P}^{c\ssc L} &=&   p^c -   i \frac{1}{k_x k_p} \partial_{x^c}\longrightarrow p^c\;,
\eea
and the Moyal star-product reduces to a simple 
commutative product. Functions $\za(p\star,x\star)$,
representing quantum observables, reduce to 
multiplicative operators  $\za(p^c,x^c)$, the 
classical observables acting on the contracted 
representation space of the original pure and 
mixed states.

For the Krein space of pure states, the coherent
state basis is taken with the new labels as 
$\left|{p}^c,{x}^c\rra$, where  $2{p}_\mu^c$ 
and $2{x}_\mu^c$ characterize the expectation 
values of $\hat{X}_\mu^c$ and $\hat{P}_\mu^c$ 
operators. We have
\bea
 \prescript{}{_\eta}{\!\lla {p}'^c_\mu,{x}'^c_\mu | \hat{X}_\mu^c |{p}_\mu^c,{x}_\mu^c\rra}
&=& [({x}'^c_\mu+{x}^c_\mu)-i\frac{k_p}{k_x}({p}'^c_\mu-{p}^c_\mu)]
 \prescript{}{_\eta}{\!\lla  {p}'^c_\mu,{x}'^c_\mu \right.  \left|{p}_\mu^c,{x}_\mu^c\rra }\;,
\nonumber \\
 \prescript{}{_\eta}{\!\lla  {p}'^c_\mu,{x}'^c_\mu | \hat{P}_\mu^c |{p}_\mu^c,{x}_\mu^c\rra}
&=& [({p}'^c_\mu+{p}^c_\mu)+i \frac{k_x}{k_p} ({x}'^c_\mu-{x}^c_\mu)]
 \prescript{}{_\eta}{\!\lla  {p}'^c_\mu,{x}'^c_\mu \right.  \left|{p}_\mu^c,{x}_\mu^c\rra} \;,
\eea
with $ \prescript{}{_\eta}{\!\lla  {p}'^c_\mu,{x}'^c_\mu \right.  \left|{p}_\mu^c,{x}_\mu^c\rra}$  
at the contraction limit going to zero for two 
distinct states. Note that the  ${k_p}$-${k_x}$ 
ratio is, at the contraction limit, a constant with 
physical dimension and it is showing up in the 
above equations only to take care of the 
difference in physical units for $p^c$ and $x^c$. 
The Krein space, as a representation for the 
contracted symmetry, as well as a representation 
of the now commutative algebra of observables, 
reduces to a direct sum of one-dimensional 
representations of the ray spaces of each 
$\left|{p}_\mu^c,{x}_\mu^c\rra$. The only 
admissible pure states are the exact coherent 
states, and not any linear combinations.  The 
obtained coherent states can be identified as 
classical states, on the space of which the 
$\tilde{G}_{\!s}$-type operators act as 
generators of symmetries.  ${G}_{\!s}\star$-type 
operators, as  general $\za\star$ in the original 
observable algebra, contract to commuting 
multiplicative operators corresponding to classical 
observables. Results suggest that the projective 
Krein space, the true quantum phase space, in 
classical limit gives exactly the classical phase space 
with ${p}_\mu^c$ and ${x}_\mu^c$ coordinates. 
The Krein space, or Schr\"odinger, picture at the 
classical limit serves rather as the Koopman-von 
Neumann formulation in a broader setting of mixed 
state, {\em i.e.} statistical mechanics. We do not 
intend to explore that aspect further in this article. 
The observable algebra, or Heisenberg picture, 
gives a much more direct way of examining the 
full dynamical theory at that contraction limit. 
It also gives a direct and intuitive picture of the 
phase space geometry too. The original position 
and momentum operators, ${x}_\mu\star$ and 
${p}_\mu\star$, can be seen as noncommutative 
coordinates of the noncommutative symplectic 
geometry of the phase space \cite{078}. The 
contracted versions as ${x}_\mu^c$ and 
${p}_\mu^c$ are the classical phase space 
coordinates with no noncommutativity left.

Let us turn to the noncommutative Hamiltonian 
transformations. As mentioned above, at the 
quantum level, a ${G}_{\!s}\star={G}_{\!s}({p}_\mu\star, {x}_\mu\star)$
operator is a Hamiltonian function of the phase 
space coordinates $p\star$ and $x\star$, and the 
corresponding $\frac{-1}{2i}\tilde{G}_{\!s}$ is the 
Hamiltonian vector field. It is, of course, well known 
since Dirac that what has now been identified 
as a quantum Poisson bracket $\frac{1}{2i}[\cdot,\cdot]$
 \cite{078,081} (and see references therein) reduces 
exactly to a classical Poisson bracket, which works 
in our formulation, explicitly shown in Ref.\cite{070}; 
{\em i.e.}
\[
 {G}_{\!s}({p}_\mu\star, {x}_\mu\star)  \to {G}_{\!s}^c({p}_\mu^c,{x}_\mu^c)\;,
\qquad
 \frac{-1}{2i}\tilde{G}_{\!s}=\frac{1}{2i}[\cdot,\cdot]
\to \{\cdot,{G}_{\!s}^c\}= \frac{-1}{2i}\tilde{G}_{\!s}^c \;.
\]
The explicit expressions are in exactly the same 
form as those of the quantum case, namely
\bea &&
\tilde{G}_{\!\omega^{\mu\nu}}^c= \tilde{G}_{\!\omega^{\mu\nu}}
=- 2i(x_\mu^c \partial_{x^{c\nu}} -p_\nu^c \partial_{p^{c\mu}}) 
    -(\mu \leftrightarrow \nu) \;,
\sea 
\tilde{G}_{\!-x^{c\mu}} =-2i \partial_{x^{c\mu}} \;, 
\qquad\qquad   
\tilde{G}_{\!p^{c\mu}} = 2i \partial_{p^{c\mu}} \;.
\eea
Note their independence on the contraction 
parameter $k$ (or $k_p$ and $k_x$), even before 
the $k \to \infty$ limit is explicitly taken. In 
conclusion, from the quantum Poisson bracket 
in terms of the Moyal bracket, or the Hamiltonian 
vector field given in terms of $\tilde{G}_{\!s}$,
we retrieve the Hamiltonian flow equation
\bea
\frac{d}{ds} \za(p^c,x^c) = \{ \za(p^c,x^c), {G}_{\!s}^c\} = \frac{-1}{2i}\tilde{G}_{\!s}^c \za(p^c,x^c) \;
\eea 
for any classical observable $\za(p^c,x^c)$ as a function 
of basic observables $x^{c\mu}$ and $p^{c\mu}$, which 
also serve as canonical coordinates for the phase space, 
with the standard expression for the classical Poisson 
bracket. The Hamilton's equations (\ref{heq}), as 
specific example, become
\bea
\frac{d}{d\tau}x^c_\mu =\frac{\partial G_{\tau}^c}{\partial p^{c\mu}} 
  = \frac{p^c_\mu}{m}
\qquad \qquad
\frac{d}{d\tau}p^c_\mu =-\frac{\partial G_{\tau}^c}{\partial x^{c\mu}}
  = -\frac{\partial v(x^{c\nu})}{\partial x^{c\mu}}\;.
\eea
$G_{\tau}^c=\frac{p^{c\mu}p^c_\mu}{2m}$
is the covariant classical Hamiltonian.

\section{Conclusions}

We presented  a formulation of covariant quantum
mechanics as an irreducible component of the regular 
representation of  the $H_{\!\ssc R}(1,3)$ (quantum) 
relativity symmetry, with a pseudo-unitary inner product 
essentially obtained from an earlier study of the covariant 
harmonic oscillator problem identified as a representation 
of the same symmetry \cite{083}. The pseudo-Hermitian
nature of operators in the observable algebra is emphasized, 
with a metric operator $\hat\eta$ as the exact quantum 
manifestation of the Minkowski metric for the classical 
spacetime. The natural wavefunction representation 
$\phi(p^\mu,x^\mu)$ is the one in a coherent state basis. 
The Fock states as eigenstate solutions to the covariant 
harmonic oscillator Hamiltonian are a great orthonormal 
basis for the Krein space as a representation
space. Actually, the overcomplete set of coherent
states and the position and momentum operators as
differential operators all have the usual form exactly
as in the otherwise unitary representation, completely
hiding the incompatibility of the latter with the Fock 
state system assuming an invariant $n=0$ state. That
seems to have made the incompatibility to escape the
attention of the previous authors. Our different starting
perspective \cite{082} and a careful analysis,
especially in the language of pseudo-Hermitian
quantum mechanics allowing a general metric 
operator $\hat\eta$, and hence a general metric/inner 
product  on the space of state vectors, illustrate the 
proper mathematical description well. In particular,
we obtain explicit form of the nondegenerate but 
indefinite inner product for the wavefunctions 
$\phi(p^\mu,x^\mu)$, with a nontrivial integration
measure, to go along with the $\eta$-Hermitian nature 
of the position and momentum operators. Though the 
wavefunctions for the Fock states are divergent at 
timelike infinity, the `probability amplitude' is finite
over any parameter interval. As a complete solution
to the covariant harmonic oscillator problem, our 
results have all the desirable features which have
not otherwise been fully available.

To retrieve the standard probability interpretation for 
the formulation of Lorentz covariant quantum mechanics,
one can simply project the theory onto the Lorentz
invariant subspace of positively normed states.
The very nice properties of the full theory under the 
Lorentz transformations, again  well illustrated in
terms of the Fock  basis, assure the projection
does not lead to any undesirable feature. 

Our study is a part of our fully quantum relativity 
group-theoretically based program. The constructed 
quantum mechanics is the `relativistic' version of 
the so-called `non-relativistic' theory based on the 
$H_{\!\ssc R}(3)$ group, or on the $\tilde{G}(3)$ group,  
a $U(1)$ central extension  of the Galilean group.  
$H_{\!\ssc R}(3)$ is a subgroup of the $H_{\!\ssc R}(1,3)$ 
group, while together with $\tilde{G}(3)$ they are both 
subgroups of the $c \to \infty$ approximation of the 
$H_{\!\ssc R}(1,3)$, obtained as a symmetry, or Lie 
algebra, contraction. The study here successfully
completes the full picture from `relativistic' quantum 
mechanics down to the `relativistic' classical and
the `non-relativistic' quantum and classical theories
as successive contractions/approximations.

Some comments on the relation of the work to the
noncommutative geometric perspective of quantum
physics and quantum spacetime may be in order.
Our idea of having the pseudo-unitary metric on the 
space of states comes mostly from the intuition on 
the need to take the pseudo-unitary Minkowski metric 
seriously as a quantum notion, for the noncommutative 
position and momentum operators as coordinates for 
the space \cite{082}. The quantum phase space, exactly 
the projective Hilbert space for the `non-relativistic' 
theory has been shown to serve as the quantum model 
of the physical space \cite{066,070}. Well known 
as an infinite dimensional symplectic manifold, 
a noncommutative  geometric picture of it has been
presented \cite{078} with the position and momentum 
operators as coordinates. A new conceptual notion of
noncommutative values for the quantum observables
 \cite{079} has been introduced to achieve a consistent
interpretation of the values of the six coordinates 
being able to specify a point in the phase space otherwise
described by an infinite number of real number coordinates.
The noncommutative value of an observable for a state
carries the full information the mathematical formulation 
of the theory actually contains for that. The notion gives
an intuitive, but noncommutative, picture of quantum
physics \cite{081}; perhaps also a noncommutative 
notion of `local realism'. The current Lorentz covariant 
theory is what would allow an analogous picture with 
$\hat{X}^\mu$ and $\hat{P}^\mu$ as coordinates 
bearing the Minkowski nature. Our background
group theoretical framework has a stable symmetry
with $X$-$X$ and $P$-$P$ type noncommutativity
to which the $H_{\!\ssc R}(1,3)$ symmetry is a 
contraction limit  \cite{030,071}. Other forms of
covariant $X$-$X$ noncommutative geometry 
with or without consideration of gravitation have
been available in the literature(see for examples
Refs.\cite{31,32,33}). We share the belief that the 
proper theory of quantum gravity has to be a
geometrodynamics of quantum, noncommutative,
spacetime. Our framework has the unique feature 
that one has to take the `phase space' as the model 
for the spacetime, as one irreducible representation.

The work focuses only on the formulation aspects, 
establishing such a theory that has all the nice properties
mentioned and can successfully address the various 
concerns raised on such covariant theories. It may be
considered mathematically involved, but unfortunately
quite necessary. The only practical physical problem we 
have addressed is the free particle case, and arguably the
covariant harmonic oscillator problem, with the latter
being of great theoretical importance. However, practical
application of `relativistic' quantum mechanics is 
generally tricky \cite{BD}.  Even with dynamics for the 
electron, quantum field theory is usually preferred. 
Applying the usual Dirac equation at the particle dynamics 
level has to confront the tricky issue of the negative energy 
solutions and the related {\em zitterbewegung}.  There
is also the fact that it gives 
$\frac{d\hat{X}^i}{dt} = - \frac{c}{2}  \gamma^{\ssc 0}\gamma^i \ne \frac{\hat{P}^i}{m}$,
(here, $\gamma^{\ssc 0}=\mbox{\tiny $\left( \begin{array}{cc} 0 & -1 \\-1 & 0 \end{array} \right)$}$
and $2$ for $\hbar$).  
Otherwise, the `nonrelativistic' and classical limits 
of the theory are well analyzed (see for examples 
Refs.\cite{sc,Ba}).  On the other hand, our group 
theoretical framework naturally gives the dynamical 
theories in a covariant symplectic formalism with 
dynamical evolution to be described through an 
invariant parameter, like $\tau$. The equation of
motions, spin zero and probably also for the higher
spin cases, are really in the form initiated by
St\"uckelberg early the 1940's (see Refs.\cite{H,Fc}
and references therein). The studies stick with 
unitary representations of the Poincar\'e symmetry,
though having to live with  a non-Hermitian 
$\gamma^\mu \hat{P}_\mu$ for a Dirac spinor. 
Yet there are some nice theoretical features from 
the theory, including  good position operators
$\hat{X}^\mu$. Modifying the spin $\frac{1}{2}$ 
theory to our pseudo-Hermitian setting has a good 
chance of improving its physics picture better. We 
plan on taking the theory more seriously along the 
line of our $H_{\!\ssc R}(1,3)$ representation 
framework in  terms the formulation before going 
into studies of  practical systems. Hope to report 
on the results in the near future.

\bigskip\bigskip\noindent
\textbf{Appendix  : Illustration of problems in unitary 
formulation of covariant harmonic oscillator}

In this appendix, we summarize the standard 
approach to covariant harmonic oscillator problem, 
which attempts to construct a unitary Fock space,
assuming the position and momentum operators
being Hermitian. A special attention will be given to 
its problems \cite{Z,B}, here especially as seen in 
the ${\phi}\!(p^\mu,x^\mu)$ wavefunction 
picture, which are all avoided  in our 
pseudo-unitary representation. The difference in 
the two representations is in the inner product, 
which is simply
\[
 \lla \phi | \phi' \rra  =\frac{1}{\pi^4}  
 \int\!\! d^4p \,d^4x \; 
     \bar{\phi}(p^\mu,
     x^\mu) \, {\phi'}\!(p^\mu,x^\mu) \;,
\]
for the unitary case. The first sign of the problem 
arises already in the abstract vector space.

The ladder operators are given essentially in 
the same way,  $\hat{a}^\mu=\eta^{\mu\nu} (\hat{X}_\nu+i\hat{P}_\nu)$,
$\hat{a}^{\dag}_\mu=\hat{X}_\mu-i\hat{P}_\mu$, 
where we drop the corresponding trivial $\hat\eta$.
As illustrated in the main text, the abstract algebraic
analysis is not  sensitive to the nature of
the metric $\hat\eta$. The same conclusion of
$\lla m| n \rra=(-1)^{n_0} \delta_{mn}$ cannot
be avoided. \cite{B}. So, the Fock space is still 
the same Krein space, which then cannot be the 
Hilbert space of the unitary representation of
the $H_{\!\ssc R}(1,3)$ symmetry. The only way
to avoid that is to take a $|0\rangle$ state that is 
not Lorentz invariant \cite{B}, meaning that the 
Lorentz symmetry is spontaneously broken in 
the system, which hardly sounds like the quantum
version of the classical covariant harmonic oscillator 
system or anything we may have a good reason to 
be interested in. The key thing is that the noncompact 
nature of $SO(1,3)$ gives no finite-dimensional 
unitary representation. Since the Hamiltonian for 
the problem, or the operator $\hat{N}$, is Lorentz 
invariant, the $n$-level subspaces are likewise 
invariant and hence can only be infinite-dimensional, 
so long as unitary representations
are concerned. The states on a fixed $n$-level 
do not transform as symmetric Minkowski 
$n$-tensors. That is the key issue behind the
incompatibility of the latter and the kind of nice 
physics picture one would like to have for the
system  \cite{082}, which our pseudo-unitary 
formulation successfully retrieved.

Now, let us turn to the wavefunction representation.
The Fock state wavefunctions are eigenfunctions of 
\[
\hat{N}=\frac{1}{4}\left( x_\mu x^\mu+p_\mu p^\mu-\frac{\partial^2}{\partial p_\mu \partial p^\mu}
    -\frac{\partial^2}{\partial x_\mu \partial x^\mu} +2 i x^\mu \partial_{p^\mu}
        -2 i p^\mu \partial_{x^\mu} \right) -2
\]
operator. One can easily check that
\[
     \phi_n\left(p^\mu,x^\mu\right) 
= e^{-\frac{x_\mu x^\mu+p_\mu p^\mu}{2}}\prod_{\mu=0}^3\left(x_\mu-i p_\mu\right)^{n_\mu}
\]
are  solutions for the eigenvalue 
$n_{ \ssc 0}+n_{ \ssc 1}+n_{ \ssc 2}+n_{ \ssc 3}$,
with $\phi_o$ corresponding to $|0 \rangle$ state. To stick
to the probability interpretation with the trivial
measure in the integral inner product, one has to
restrict the domain of the wavefunctions to 
spacelike region of $p^\mu$ and $x^\mu$.

In order to normalize the wavefunction, we 
need to calculate the integral 
\[
   \int \frac{d^4x d^4 p}{\pi^4}\bar{\phi}_{n}
  \left(p^\mu,x^\mu\right) \phi_{n} \left(p^\mu,x^\mu\right)
=\int \frac{d^4x}{\pi^2} 
  \int \frac{d^4 p}{\pi^2} e^{-x_\mu x^\mu} e^{-p_\mu p^\mu}
    \prod_{\mu=0}^3 \left(x_\mu^2 +p_\mu^2\right)^{\!n_\mu}
\]
over the parameter domain. Without the domain
restriction the integral surely diverges. Let us focus on the 
parts of the integral for the $\prod_{\mu=0}^3 x_\mu^{2n_\mu}$
term, and similarly the  $\prod_{\mu=0}^3 p_\mu^{2n_\mu}$ 
term, from the expansion of the last factor. 
Specifically, we have an integral of the form
$  \int \frac{d^4 x}{\pi^2} e^{-x_\mu x^\mu} \prod_{\mu=0}^3 x_\mu^{2n_\mu}$
to deal with. The  integral can be evaluated with 
coordinates in a polar form as in Ref.\cite{Z},  by defining 
$r^2=x_\mu x^\mu$, $\rho=x^0/\sqrt{x_i x^i}$, 
and the spatial angular coordinate  of 
which we skip the details. We obtain
\begin{equation}
    \mathcal{I}_{\!\ssc 0}
  =\frac{1}{\pi^2}\int d\Omega \int_0^\infty dr\,r^3 e^{-r^2}\int_{-1}^1\frac{d\rho}{(1-\rho^2)^2}
   =\frac{2}{\pi}\int_{-1}^1\frac{d\rho}{(1-\rho^2)^2}
\end{equation}
for the $n=0$ case, with $\Omega$ denoting the
spatial solid angle. The $\rho$-integral is still divergent. 
The integrand for the specific case is $\rho$-independent. 
The divergence is simply due to the infinite range of 
the boost parameter ($\rho$ being its hyperbolic tangent).
Hence, it has been suggested to define the integral 
with `the infinite volume factor' absorbed \cite{Z,B}.
However, there is really no sensible way to do that
so long as the Fock state wavefunctions are concerned.
The corresponding $\rho$-integral of $\phi_n$ for
a nonzero $n_{\!\ssc 0}$ has an extra factor of 
$\frac{\rho^{2n_{\!\ssc 0}}}{(1-\rho^2)^n}$ from 
the integrand, giving a  higher order divergence 
for each larger $n$ value. From the structure of 
the full integral inner product , it is clear that such 
contributing terms of the higher order divergence
stay. That is to say, none of Fock state wavefunctions 
are really normalizable under the unitary formulation. 
This is not an artefact of the coherent state framework. 
The usual Schr\"odinger wavefunctions $\psi(x^\mu)$ 
for the Fock states have the same problem.

There is an alternative approach of taking a timelike,
instead of spacelike, parameter restriction for the 
domain of the wavefunctions, really corresponding
to defining $|0 \rangle$ as satisfying 
$a^\dag |0 \rangle =0$. Similar problems persist.

\vspace*{.2in}
\noindent{\bf Acknowledgements \ }
S.B. thanks the Center for High Energy and High Field Physics,
National Central University for hospitality.
 O.K. and H.K.T. are partially supported by research grant 
number  109-2112-M-008-016
of the MOST of Taiwan.


\begin{thebibliography}{999}
\bibitem{070}
Chew, C.S.; Kong, O.C.W.; Payne, J.  
Observables and Dynamics Quantum to Classical from a
Relativity Symmetry and Noncommutative-Geometric Perspective.
\emph{ J. High Energy Phys. Gravit. Cosmol.}  {\bf 2019}, {\em
 5}, 553--586, doi:10.4236/jhepgc.2019.53031. 
 \bibitem{u1}
 de Azc\'arraga, J.A.;  Izquierdo,  J.M.
{\it Lie Groups, Lie Algebras, Cohomology and Some Applications in Physics};  Cambridge University Press: New York, NY, USA, 1995.
\bibitem{cs1} 
%
  {Perelomov,  A.M.}  Generalized Coherent States and Some of their Applications. \emph{Sov. Phys. Usp.} {\bf 1977}, {\it 
 	20},  703--720.
\bibitem{cs2} 
  {Zhang,  W.-M.;}  Feng,  D.H.;  Gilmore,   R. Coherent States: Theory and Some Applications. 
\emph{Rev. Mod. Phys.} {\bf 1990}, {\it
 62}, 867--927, doi:10.1103/RevModPhys.62.867.
\bibitem{cs} 
 Klauder,   J.R. \textit{A Modern Approach to Functional Integration};  Birkh\"auser:  Secaucus, NJ, USA, 2010. 
\bibitem{c*1}
 Palmer,    T.W. \textit{Banach Algebras and the General Theory of $^{\ast}$-Algebras Vol II}; 
 Cambridge University Press:  Cambridge, UK, 2001.
\bibitem{c*2}
  Pedersen,   G.K. \textit{$C^*$-algebras and their Automorphism Groups}; Academic Press:  London, UK, 1979. 
\bibitem{M} 
  Dubin,  D.A.;  Hennings,  M.A.;  Smith,  T.B.
{\em Mathematical Aspects of Weyl Quantization and Phase};  World Scientific: Singapore, 2000.
\bibitem{w1} 
 %
 {Hansen, F.  Quantum Mechanics in Phase Space. \emph{Rep. Math. Phys.} {\bf 1984}, {\it
 19},  361--381, doi:10.1016/0034-4877(84)90008-9. }
\bibitem{w2}
  Gracia-Bond\'ia, J.M.;   V\'arilly, J.C. Algebras of Distributions Suitable for Phase-space Quantum Mechanics I.  \emph{\mbox{J. Math. Phys.}} {\bf 1988}, {\it
 29},  869--879, doi:10.1063/1.528200. 
\bibitem{w3}
   Zachos,   C.K.;  Fairlie,    D.B.;  Curtright,  T.L.
\textit{Quantum Mechanics in Phase Space: An Overview with Selected Papers}; World Scientific: Singapore, 2005. 
\bibitem{078}
 Kong,  O.C.W.;  Liu, W.-Y. Noncommutative Coordinate Picture of the Quantum Phase Space. 
{\em arXiv: 1903.11962, NCU-HEP-k078} {\bf 2019}.
\bibitem{081}
   Kong, O.C.W. A Geometric Picture of Quantum 
Mechanics with Noncommutative Values for Observables. 
\emph{Results  Phys.} \textbf{2020}, \emph{19}, 103606,
doi:10.1016/j.rinp.2020.103636.
\bibitem{066} 
  Chew,   C.S.; Kong,  O.C.W.;  Payne,  J.
 A Quantum Space Behind Simple Quantum Mechanics.
\emph{Adv. High Energy Phys.} {\bf 2017}, {\em
 2017}, 4395918. doi:10.1155/2017/4395918.
\bibitem{BL}
  Bacry,  H.;  L\'evy-Leblond,  J.-M. Possible Kinematics.
\emph{J. Math. Phys.} {\bf 1968}, {\em
 9},  1605--1614, doi:10.1063/1.1664490.
\bibitem{030}
   Kong, O.C.W. A deformed relativity with the quantum $\hbar$. 
\emph{Phys. Lett. B}  {\bf 2008}, {\em
 665},  58--61, doi:10.1016/j.physletb.2008.05.060.
\bibitem{071}
   Kong,  O.C.W.;  Payne,  J.
The First Physics Picture of 
Contractions from a Fundamental Quantum Relativity Symmetry 
Including All Known Relativity Symmetries, Classical and Quantum. 
 \emph{Int. J. Theor. Phys.}  {\bf 2019}, {\em
 58},  1803--1827, doi:10.1007/s10773-019-04075-x.
\bibitem{Z} 
  Zmuidzinas,  J.S. Unitary Representations of the Lorentz Group on 4-Vector Manifolds.
 \emph{J. Math. Phys.}  {\bf 1966}, {\em
 7}, 764--780, doi:10.1063/1.1704991.
\bibitem{J}
   Johnson,   J.E. Position Operators and Proper Time in Relativistic Quantum Mechanics.
 \emph{Phys. Rev.}  {\bf 1969},  {\em
  181},  1755--1764, doi:10.1103/PhysRev.181.1755. 
\bibitem{B} 
 Bars,  I.  Relativistic Harmonic Oscillator Revisited. 
\emph{Phys. Rev. D} {\bf 2009}, {\em
 79}, 045009, doi:10.1103/PhysRevD.79.045009.
\bibitem{082}
  Kong,  O.C.W. 
The Case for a Quantum Theory on a Hilbert Space with an Inner Product of
Indefinite Signature. 
\emph{J. High Energy Phys. Gravit.  Cosmol.} {\bf 2020}, {\em
  6},  43--48, doi:10.4236/jhepgc.2020.61005.
\bibitem{083}
 Bedi\'c,  S.;  Kong,  O.C.W. 
Analysis on Complete Set of Fock States with Explicit Wavefunctions for the Covariant Harmonic Oscillator Problem. 
\emph{Symmetry} {\bf 2020}, {\em
 12},  39,  doi:10.3390/sym12010039.
\bibitem{D}
  Dirac,  P.A.M. The Physical Interpretation of Quantum Mechanics. 
 \emph{Proc. R. Soc. Lond. A} {\bf 1942}, {\em
 180},  1--40, doi:10.1098/rspa.1942.0023.
\bibitem{P}
  Pauli,   W. On Dirac's New Method of Field Quantization. 
\emph{Rev. Mod. Phys.} {\bf 1943}, {\em
 15}, 175--207, doi:10.1103/RevModPhys.15.175.
\bibitem{N} 
  Nagy,  K.L. {\em State Vector Spaces with Indefinite Metric in Quantum Field Theory}; 
P. Noordhoff Ltd.: Groningen, The Netherlands, 1966. 
\bibitem{GB1}
  Gupta, S.N.
Theory of Longitudinal Photons in Quantum Electrodynamics. 
 \emph{Proc. Phys. Soc. A} {\bf 1950},  {\em
 63},  681--691, doi:10.1088/0370-1298/63/7/301.  
\bibitem{GB2}
 Bleuler,   K.
 Eine neue Methode zur Behandlung der longitudinalen und skalaren Photonen. 
 \emph{Helv. Phys. Acta} {\bf 1950}, {\em
 23}, 567--586.
\bibitem{Bo}
 Bogn\'ar,    J. 
 {\it Indefinite Inner Product Spaces}; Springer: Berlin, Germany, 1974.
\bibitem{pH1}
 Bender,    C.M. Introduction to PT-Symmetric Quantum Theory. 
\emph{Contemp. Phys.} {\bf 2005}, {\em
 46}, 277--292, doi:10.1080/00107500072632.
\bibitem{pH2}
 Das, A. 
 Pseudo-Hermitian Quantum Mechanics.
\emph{J. Phys. Conf. Ser.} {\bf 2011}, {\em
 287}, 012002, doi:10.1088/1742-6596/287/1/012002.
\bibitem{pH-M}
  Mostafazadeh,  A.
Pseudo-Hermitian Representation of Quantum Mechanics.
 \emph{Int. J. Geometr. Methods Modern~Phys.} {\bf 2010}, {\em 
 7}, 1191--1306, doi: 10.1142/S0219887810004816.
\bibitem{T}
  Taylor,   M.E. {\it Noncommutative Harmonic Analysis}; American Mathematical Society:  {Providence, RI, USA}, 1986.
\bibitem{BR} 
 Barut, A.O.;   Raczka,  R. {\it Theory of Group Representations and Applications}, 2nd rev. ed.; 
Polish Scientific Publishers: Warsaw, Poland, 1980. 
\bibitem{G}
 Gilmore, R.   {\it Lie Groups, Lie Algebras, and Some of Their Applications}; {Dover Publications, Inc.: Mineola, New York, USA}, 2005.
\bibitem{IW}
  In\"on\"u, E.;   Wigner,  E.P.
On the Contraction of Groups and their Representations.
 \emph{Proc. Natl. Acad. Sci. USA} {\bf 1953}, {\em 39}, 510--524, doi: 10.1073/pnas.39.6.510.
%
\bibitem{060}
  Cho,  D.-N.;  Kong,    O.C.W.  Relativity Symmetries and Lie Algebra Contractions. 
  \emph{Ann. Phys.}  {\bf 2014}, {\em 351}, 275--289, doi:10.1016/j.aop.2014.09.005.
\bibitem{079}
  Kong,  O.C.W.;  Liu,  W.-Y. Noncommutative  Values of Quantum Observables. 
 \emph{Chin. J. Phys.}  \textbf{2021}, {\it 69}, 70--76,
doi:10.1016/j.cjph.2020.11.008
\bibitem{31}
{Steinacker, H. Emergent gravity on covariant quantum spaces in the IKKT model. 
\emph{J. High Energy Phys.} {\bf 2016}, 156.}
\bibitem{32}
 Sperling, M.;  Steinacker, H. Higher spin gauge theory on fuzzy $S^4_N$. 
\emph{J. Phys. A  Math. Theor. }   \textbf{2018}, \emph{51},  075201.
\bibitem{33}
Bojowald, M.; Brahma, S.;  Buyukcama, U;   Ronco, M.
Extending general covariance: Moyal-type noncommutative manifolds. 
\emph{Phys. Rev. D} \textbf{2018}, \emph{98}, 026031.
\bibitem{BD}
 Bjorken, J.D.;   Drell, S.D.
{\em Relativistic Quantum Mechanics}; {McGraw-Hill Inc.: New York, USA}, 1964. 
\bibitem{sc}
 {Mosses, R.E.} The Reduction of the Dirac Equation to a Nonrelativistic Form. \emph{Am. J. Phys.}  \textbf{1971}, \emph{39}, 1169--1172. 
\bibitem{Ba}
Baylis, W.E.  Classical eigenspinors and the Dirac equation. \emph{Phys. Rev. A}  \textbf{1995}, \emph{45},  4293--4302. 
%
\bibitem{H}
{Horwitz, L.P. {\em Relativistic Quantum Mechanics}; 
Springer Science+Business Media: Dordrecht, The Netherlands, 2015.}
\bibitem{Fc}
  Fanchi, J.R. {\em Parametrized Relativistic Quantum Theory}; 
Kluwer Academic Publishers:  Dordrecht, The~Netherlands,  1993.
\end{thebibliography}
\end{document}